\documentclass{aa}  

\usepackage{graphicx}
\usepackage{multirow}
\usepackage{xcolor}
\usepackage[caption=false]{subfig}
\usepackage{bm}
\usepackage{soul}  % for strikethrough text with \st

\usepackage{txfonts}
\newcommand{\allers}{AL13}

\begin{document}

   \title{Youth analysis of near infrared spectra of young low-mass stars and brown dwarfs}
    %\titlerunning{}
    \authorrunning{Almendros-Abad et al.}
    \titlerunning{Youth analysis of near infrared spectra of young low-mass stars and brown dwarfs}
    
   \author{ V. Almendros-Abad \inst{1}, K. Mu\v{z}i\'c\inst{1}, A. Moitinho\inst{1}, A. Krone-Martins\inst{2,1} 
          \and K. Kubiak\inst{1}
          %\fnmsep
          %\thanks{Just to show the usage
          %of the elements in the author field}
          }

   \institute{CENTRA, Faculdade de Ci\^{e}ncias, Universidade de Lisboa, Ed. C8, Campo Grande, P-1749-016 Lisboa, Portugal\\
              \email{valmendros@sim.ul.pt}
         \and
            Donald Bren School of Information and Computer Sciences, University of California, Irvine, CA 92697, USA\\
            }

   \date{Received; accepted}

% \abstract{}{}{}{}{} 
% 5 {} token are mandatory
  \abstract
   {
Studies of the low-mass population statistics in young clusters are the foundation for our understanding of the formation of low-mass stars and brown dwarfs. Robust low-mass populations can be obtained through near-infrared spectroscopy, which provides confirmation of the cool and young nature of member candidates. However, the spectroscopic analysis of these objects is often not performed in a uniform manner, and the assessment of youth generally relies on the visual inspection of youth features whose behavior is not well understood. 

    }
   {We aim at building a method that efficiently identifies young low-mass stars and brown dwarfs from low-resolution near-infrared spectra, by studying gravity-sensitive features and their evolution with age.
   }
   {We built a dataset composed of all publicly available ($\sim$2800) near-infrared spectra of dwarfs with spectral types between M0 and L3. First, we investigate methods for the derivation of the spectral type and extinction using comparison to spectral templates, and various spectral indices. Then, we examine gravity-sensitive spectral indices and apply machine learning methods, in order to efficiently separate young ($\lesssim$10 Myr) objects from the field.
   }
   {Using a set of six spectral indices for spectral typing, including two newly defined ones (TLI-J and TLI-K), we are able to achieve a precision below 1 spectral subtype across the entire spectral type range. We define a new gravity-sensitive spectral index (TLI-g) that consistently separates young from field objects, showing a performance superior to other indices from the literature. Even better separation between the two classes can be achieved through machine learning methods which use the entire NIR spectra as an input.
   Moreover, we show that the H- and K-bands alone are enough for this purpose. Finally, we evaluate the relative importance of different spectral regions for gravity classification as returned by the machine learning models. We find that the H-band broad-band shape is the most relevant feature, followed by the FeH absorption bands at 1.2 $\mu m$ and 1.24 $\mu m$ and the KI doublet at 1.24 $\mu m$.
   }
   {}


   \maketitle
%
%-------------------------------------------------------------------

\section{Introduction}

Star Forming Regions (SFRs) are perfect laboratories for studying the direct outcome of star formation. The distribution of masses at birth or Initial Mass Function (IMF) provides an important constraint for testing the different theoretical scenarios of star formation \citep{kroupa01,chabrier03,bastian10,offner14}. The low-mass end of the IMF is populated by low-mass stars and brown dwarfs (BDs), whose masses can extend well below the deuterium burning limit. The formation of BDs is still a subject of debate, as very high densities are required for such a low-mass cloud fragment to follow gravitational collapse, along with a small mass reservoir to maintain them below the substellar limit \citep{padoan04,stamatellos09,bate12,whitworth04}. The main proposed mechanisms for the formation of these objects are the direct collapse of a cloud fragment with a mass below the substellar limit \citep{padoan04,whitworth04}, and disk fragmentation with a later ejection from the system \citep{basu12,bate12,thies10,stamatellos09}. The shape of the low-mass side of the IMF is difficult to constrain, as these objects are very faint and still embedded in their parent clouds. Studies in nearby star forming regions, as well as in more massive and dense environments find that the IMF seems to be universal within the observational errors, with 2 to 5 stars being born for every BD \citep{luhman07,bayo11,pena12,alvesdeoliveira12,lodieu13,scholz13,drass16,muzic17,muzic19}.

In order to have a well constrained IMF of a SFR, its member list needs to be as clean and complete as possible, ideally through spectroscopic confirmation. Photometric samples of SFRs low-mass member candidates contain not only bona-fide members, but also a collection of different kinds of contaminants, with the most significant contribution by reddened background objects. In order for a candidate to be spectroscopically confirmed as a low-mass member, its spectrum needs to show the characteristic low-mass spectral shape and display signatures of youth. The main difference between two cool dwarfs of the same mass, but different age, is their surface gravity. Very young cool dwarfs are still contracting, therefore they have a lower pressure/gravity environment at their surface (larger radii for the same mass). This fundamental difference leaves an imprint in their spectra in the form of gravity-sensitive features \citep{lucas01,gorlova03,mcgovern04,allers07}. 

Youth assessment based on near-infrared (NIR) spectroscopy, is typically carried out by visual inspection of these gravity-sensitive features \citep{lucas06,allers07,lodieu08,alvesdeoliveira12,muzic15,luhman16, zapatero17}. The first classification of cool dwarfs based on their surface gravity was presented in \citet{kirkpatrick05} and further explored in \citet{kirkpatrick06} and \citet{cruz09}. They defined four different gravity classes: $\delta$, $\gamma$, $\beta$ and $\alpha$, associated with ages $\sim$1, 10, 100 and 1000 Myr, respectively. This analysis is based on the visual assessment of gravity-sensitive features, namely weakness of alkali lines and strength of CaH and VO bands in the optical and NIR (J-band).  Some attempts have been made in order to quantify the correlation between the objects' age and its spectral shape (e.g. \citealt{canty13}), with the main drawback being the small sample sizes, which restricts the spectral type range of applicability and the precision of the classification system.

The first more substantial quantitative youth classification method was presented in (\citealt{allers13}; hereafter \allers), using low-resolution NIR spectroscopy. Here, they classify objects into three gravity classes (field, FLD-G; intermediate, INT-G; and very low gravity, VL-G) based on four different scores. The gravity classes were built consistent with the \citet{cruz09} system, the INT-G and VL-G being analogous to the $\beta$ ($\sim$100 Myr) and $\gamma$ ($\sim$10 Myr) classes respectively. Three of these scores are based on spectral indices tracing the strength of the KI doublet at 1.24 $\mu m$ and the VO and FeH molecular bands in the Z and J bands. The fourth score is based on the H-band broadband shape. This classification system was built using objects from the TWA Hydrae young moving group, objects with $\beta$ and $\gamma$ gravity optical classifications, objects with low gravity features, dusty objects, young companions, and field objects. While these methods focus on objects with intermediate ages, the main interest of this work lies in the distinction between young ($\lesssim$10Myr) and field objects, and also in how well we can differentiate SFR members from slightly older Nearby Young Moving Groups (NYMGs) objects. The upcoming multi-object spectrograph facilities (e.g. MOONS/VLT, NIRSPEC and NIRISS/JWST) will be producing large spectral datasets, requiring efficient analysis methods  for disentangling young objects from the contaminants. 
In this paper, we study the gravity-sensitive features in the range spanned by the JHK NIR bands (1.1-2.3 $\mu m$). We first build a dataset composed of all available NIR spectra of cool dwarfs (Section~\ref{dataset}). Then, we derive the spectral type (SpT) and extinction for the entire dataset by comparison with spectral templates. We also test the calibration of various SpT-sensitive indices from the literature, and define a method for spectral typing using these indices along with two newly defined ones (Section~\ref{SpT-Av}). We study the applicability of spectral indices to determination of youth, and define a new index (TLI-g) which outperforms other indices defined in the literature (Section~\ref{youth}). We also evaluate the performance of youth classification with machine learning applied to the entire spectrum and compare its performance with the newly defined TLI-g gravity-sensitive index (Section~\ref{mlearning}). The summary and conclusions are presented in Section~\ref{summary}.

%%%%%%%%%%%%%%%%%%%%%%%%%%%%%%%%%%%%%%%%%%%%%%%%%%%%%%%%%%%%%%%%%%%%%%%%%%%
\section{Dataset} 
\label{dataset}

The data used in this work is a compilation of all publicly available (to our best knowledge) reduced NIR spectra of cool objects, complemented by our own observations with SINFONI/VLT (described below), that meet the following criteria: dwarf classification with spectral type (SpT) between M0 and L5, and a spectrum at least in the K-band range. Although there are very few objects with SpT later than L2 found in SFRs \citep{bayo11,lodieu18}, this spectral range is interesting to explore as growing numbers of these objects will be found in SFRs with future facilities (JWST, ELT). The final SpT range M0-L5 includes both low-mass stars and brown dwarfs (BDs). We jointly refer to both these kinds of objects as cool dwarfs.

All the spectra have been taken from public online libraries, queried from Vizier \citep{vizier} or directly provided by the authors. We also include 13 new spectra from SINFONI/VLT first presented here (see Section~\ref{sinfoni}). 
 We reject the spectra that either present strong telluric contamination or have low signal to noise ratio.
 In Table~\ref{tab:dataset} we show the details of the dataset.

%When pasting the new version of the table, add:
%table to table*
%move caption to the top
%\hline \hline after begin{tabular}
%\hline before end{tabular}
%after obj 12:14:04.54 & -79:13:52.0: & & & & & & & & & 2020-02-28 & 5x300s \\
%after obj 12:22:48.48 & -74:10:20.6: & & & & & & & & & 2020-05-03 & 4x600s\\
%after ID & RA... line add:	 & (hh:mm:ss) & (dd:mm:ss) & (mag) & (mag) & (mas) & (mas\,yr$^{-1}$) & (mas\,yr$^{-1}$) &  & & (s) \\
\begin{table*}
	\caption{Dataset properties.}
	\begin{center}
        \begin{tabular}{c c c c c c c}       
          \hline\hline
            Reference & Number & Resolution & Instrument & Range & SpT & Group \\
            \hline
            Spex/Prism$^1$ & 568 & 200 & SpeX/IRTF & JHK & M0-L5 & Field \\
            Spex/IRTF$^2$ & 33 & 2000 & SpeX/IRTF & JHK & M3-L5 & Field \\
            Montreal$^3$ & 215 & 150-6000 & Various$^c$ & JHK & M4-L5 & NYMG$^b$, field \\
            \citet{dawson14} & 22 & 2000 & SpeX/IRTF & JHK & M5-L1 & USco \\
            \citet{mclean03} & 11 & 2000 & NIRSPEC/KECK-II & JHK & M6-L5 & Field \\
            \citet{bonefoy14} & 11 & 1500 & SINFONI/VLT & JHK/HK & M6-L1 & Field, NYMG$^b$, SFR$^a$ \\
            \citet{manjavacas14} & 1 & 1500-2000 & ISAAC/VLT & JHK & M9-L3 & Field, NYMG$^b$, SFR$^a$ (w/VL-G) \\
            \citet{manara13} & 21 & 3500-11300 & X-Shooter/VLT & JHK & M0-L0 & TWA, SFR$^a$ \\
            \citet{muirhead14} & 23 & 2700 & TripleSpec/Palomar & JHK & M0-M4 & Field \\
            \citet{luhman20} & 208 & 150,600 & SpeX,GNIRS & JHK & M4-L1 & USco, Sco-Cen \\
            \citet{alcala14} & 24 & 3500-11300 & X-Shooter& JHK & M1-M9 & Lupus \\
            \citet{covey10} & 12 & 200 & SpeX/IRTF & JHK & M2-M6 & B59 \\
            \citet{venuti19} & 13 & 3500-11300 & X-Shooter & JHK & M1-L0 & TWA \\
            \citet{lodieu08} & 19 & 1700 & GNIRS/Gemini & JHK & M8-L2 & USco \\
            \citet{muench07} & 17 & 200 & SpeX/IRTF & JHK & M7-L0 & IC348 \\
            \citet{allers13} & 33 & 200,2000 & SpeX/IRTF & JHK & M5-L4 & Field, NYMG \\
            \citet{esplin19} & 13 & 150,600 & SpeX,GNIRS & JHK & M3-L0 & Taurus \\
            \citet{esplin18} & 67 & 150 & SpeX/IRTF & JHK & M3-M9 & USco \\
            \citet{esplin20} & 98 & 150-3500 & Various$^d$ & JHK & M0-L1 & Rho oph \\
            \citet{luhman17} & 115 & 150,750 & SpeX/IRTF & JHK & M0-L0 & Taurus \\
            \citet{luhman16} & 235 & 150,800 & SpeX,GNIRS & JHK & M0-L0 & NGC1333, IC348 \\
            \citet{rojas-ayala12} & 133 & 2700 & TripleSpec/Palomar & K & M0-M8 & Field \\
            \citet{Terrien15} & 853 & 2000 & SpeX/IRTF & JHK & M0-M8 & NYMG$^b$, field \\
            This work & 10 & 1500 & SINFONI/VLT & JHK & M7-L2 & SFR$^a$, field \\
			\hline
		\end{tabular}
		\tablefoot{$^{a}$ from several SFRs; $^{b}$ from several NYMGs; $^c$ FIRE/MT, Flamingos-2/Gemini, GNIRS/Gemini, NIRI/Gemini, SpeX/IRTF; $^d$ CorMASS/MT, ARCoIRIS/CTIO, GNIRS/Gemini, SpeX/IRTF, Flamingos-2/Gemini.
		
		$^1$ \url{http://pono.ucsd.edu/~adam/browndwarfs/spexprism/library.html}
		
		$^2$ \url{http://irtfweb.ifa.hawaii.edu/~spex/IRTF_Spectral_Library/index.html}
		
		$^3$ \url{https://jgagneastro.com/the-montreal-spectral-library/}
	}
	\end{center}
	\label{tab:dataset}
\end{table*}

 We divide the spectra into three classes:
\begin{itemize}
    \item {\bf Young:} Objects belonging to young clusters with an age up to $\sim$10 Myr, which includes various young SFRs, along with the TWA Hydrae young moving group \citep[8-10 Myr; e.g][]{ducourant14, herczeg14, donaldson14, bell15}.
    \item {\bf Mid-gravity:} Objects belonging to NYMGs older than 10 Myr, or those classified as INT-G by \allers.
    
    \item {\bf Field:} Older objects with no youth spectroscopic features.
\end{itemize}

We decided not to use field objects (unknown age) with \allers~ VL-G classification. The VL-G class was built as an analog to the optical $\gamma$ gravity class \citep{kirkpatrick06} with ages $\sim$10 Myr \citep{allers13,gagne15,martin17}. Therefore, these objects would overlap with both our young and mid-gravity classes, but they have no known membership to any of the young Galactic structures.

In total, the dataset contains 2756 spectra, of which 906 are classified as young, 248 as mid-gravity, and 1602 as field.
The dataset is heterogeneous, it contains spectra obtained with different instruments, telescopes, observing strategies and data reductions, and the spectroscopic parameters have been derived using different methodologies. Having a large dataset is important from the statistical point of view, but at the same time we need to make sure that the heterogeneity of the dataset does not bias the results.

\subsection{SINFONI/VLT data} \label{sinfoni}
Ten spectra of late-type objects in nearby star forming regions (Chamaeleon I, $\rho$ Oph, Upper Scorpius, and Taurus), along with three spectra of VL-G objects from \citet{allers13} have been obtained using the integral-field spectrograph SINFONI at the ESO's Very Large Telescope (VLT;  \citealt{eisenhauer2003,bonnet2004}), under ESO program ID 097.C-0458. Observations were carried out in no-AO mode (field-of-view 8$\farcs0\times 8\farcs0$), with the J and H+K grisms, delivering spectra with the nominal spectral resolution R~$\sim 2000$, and R~$\sim 1500$, respectively. 
The basic data reduction steps, including the dark subtraction, division by the flat field, distortion correction, and wavelength calibration were performed using the SINFONI pipeline supplied by ESO. The extraction of the object spectra from the data cubes was performed by fitting a two-dimensional Gaussian to the median image of the source, which was then used to create a mask with the weights for extraction at different spaxels. An identical mask was used to extract the sky spectrum from an adjacent dithered exposure. In each dithered exposure, the source was placed in different detector quadrants, eliminating the necessity of observing separate sky frames. The sky subtraction was performed using the \textit{SkyCorr} tool \citep{noll14}, using the extracted object and sky spectra as an input. For each exposure, the \textit{SkyCorr} input parameters $FWHM$, $MIN\_LINE\_ DIST$ and $FLUXLIM$ were varied in order to find an optimal combination which minimises the residuals. We find that \textit{SkyCorr} performs better when $H$ and $K$ spectra are treated separately, removing the region of strong water absorption between the two bands. 

To calculate the telluric absorption spectra, we used \textit{Molecfit} \citep{smette15,kausch15}, which fits synthetic transmission spectra to astronomical data, in our case the telluric standard stars (B-type) observed immediately before or after the science observations, and at a similar airmass. The instrument response curve for each filter was determined by dividing the standard star spectra corrected for telluric absorption, with the spectrum of a black body of an appropriate effective temperature, and fitting a third degree polynomial, which helps eliminating the residual stellar features, and preserves the broad shape of the response curve. A total of 19 ($J$) and 15 ($H+K$) individual response curves observed over a course of 11 months were mean combined to produce the final response curve for each of the two filters. We find that the response curves of SINFONI in both $J$ and $H+K$ deviate by less than $4\%$ from the average one, over the period of execution.

The individual sky-subtracted spectra, corrected for telluric absorption and instrumental response, were median combined, and the noise was calculated as the standard deviation of the individual spectra at each wavelength pixel.
Finally, we scaled the $J$, $H$, and $K$ spectra according to the available photometry in individual near-infrared bands, integrating the spectra with the appropriate filter transmission curves, and doing the same on the spectrum of Vega \citep{rieke08}.

The spectra are shown in Appendix~\ref{appendix_sinfoni}.

\subsection{Giants} \label{giants}

Giants can be an important source of contamination in samples of young candidates from SFRs (e.g. \citealt{comeron13,muzic14}). They are numerous and have SpTs in the M0-M9 range. Because of their high brightness, they appear as background reddened objects simulating the population of the SFR under study. Their surface gravity is lower than that of young members of SFRs.

We compiled spectra of type III giants from the IRTF/SpeX library \citep{cushing05,rayner09}, \citet{lancon00} and the SpeX/prism library. When comparing their spectral shapes to those of dwarfs, we notice the following:
\begin{itemize}
    \item While alkali lines are as weak as in young dwarfs, FeH and CO absorption bands are stronger than for young dwarfs.
    \item The giants spectral sequence in the NIR does not resemble closely that of dwarfs, with a K-band slope that is clearly different from dwarfs of the same spectral type.
\end{itemize}

A spectroscopic distinction between giants and dwarfs, both field and young, is straightforward even at very low spectral resolution \citep{gorlova03}. We therefore exclude giants from the present analysis.

%--------------------------------------------------------------------
\section{Spectral type and extinction} \label{SpT-Av}
%--------------------------------------------------------------------

The  main  spectroscopic  property  defining  any star is the SpT, intimately related with the temperature of the object’s photosphere. However, objects that are still embedded in their parental clouds have a non-negligible extinction that needs to be taken into account, as it changes the slope of the spectra and affects the determination of the SpT.

The main method for SpT derivation is direct comparison with spectra of objects with well defined SpT (spectral templates), with the SpT being adopted from the closest match. Extinction can also be included as an additional free parameter \citep[e.g. ][]{allers07,alvesdeoliveira12,muzic14,luhman17,jose20}.
Another commonly used method for spectral type derivation are spectral indices \citep[e.g. ][]{weights09,covey10,rojas-ayala12,muzic12,allers13,alcala14}, defined as flux ratios of two or more bands that correlate well with SpT. In this case the extinction needs to be obtained through a separate method, and accounted for before calculating the SpT, unless an index is defined as extinction-independent.

In this section we study the derivation of the SpT on our dataset using these two methods: comparison with spectral templates and spectral indices. Extinction is derived by comparison with spectral templates.

%--------------------------------------------------------------------
\subsection{Comparison with spectral templates} \label{templates}
%--------------------------------------------------------------------

\begin{figure}[hbt!]
    \centering
    \includegraphics[width=\textwidth/21*10]{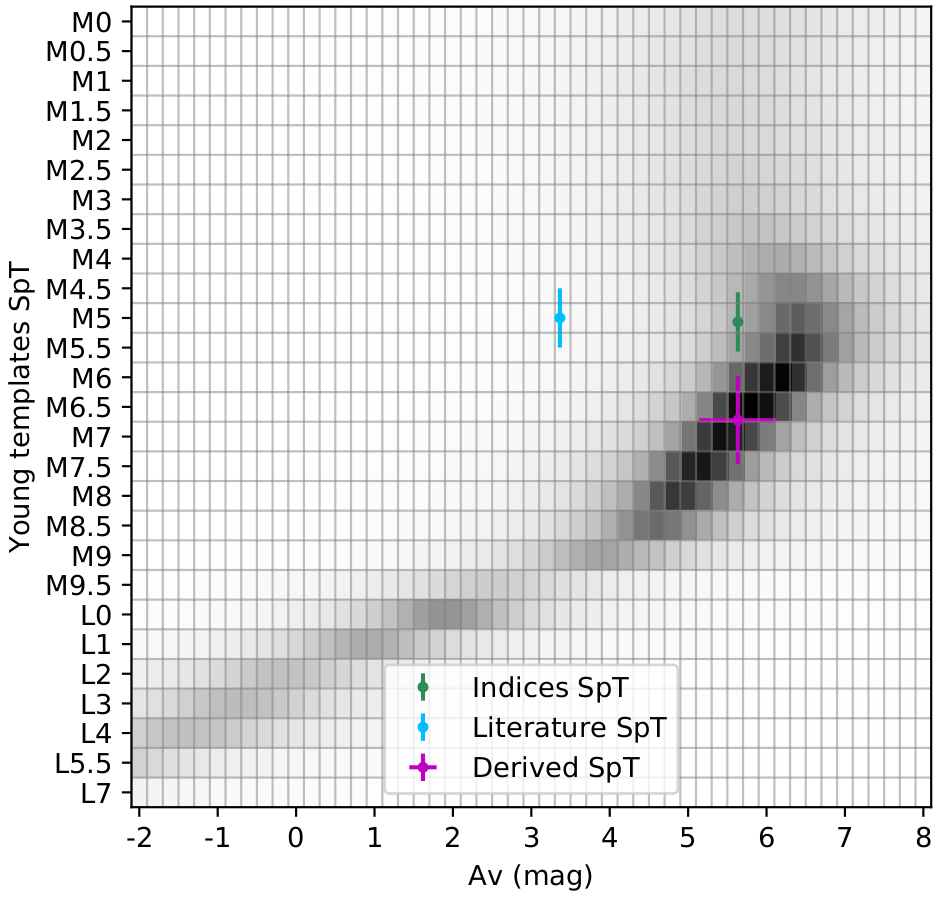}
    \caption{Distribution of normalized probabilities from the comparison with the young templates for the object Brooke18NE (known also as [BHB2007] 18NE;\citealt{covey10}). We show the literature values in cyan, the values derived using spectral templates (magenta) and the SpT derived using spectral indices (green).}
    \label{fig:probmap}
\end{figure}

We derive SpT and extinction by comparison with spectral templates. This homogeneously retrieved SpT and extinction will be the basis for the analysis performed in the following sections.
Spectral templates are built from objects showing the canonical spectral shape characteristic of their SpT, and lacking any peculiarities.
The age of an object affects its spectral energy distribution, such that comparison with field templates will systematically produce later SpTs for young objects \citep{luhman04}. We therefore need two sets of templates to characterize our dataset.
For the field sources, we use the templates from \citet{kirkpatrick10}, which are defined for integer SpTs from M0 down to L9, although we use them up to L5.
Furthermore, we use the set of young ($<$10 Myr) spectral templates built by \citet{luhman17}, which are defined for half-integer SpTs from M0 down to L0, integer SpTs for the range L0 - L4, complemented by L5.5 and L7 templates. 

We find that the young templates fit better young objects with ages up to 50 Myr, the rest of the objects are fitted with the field templates. Each spectrum in the dataset is directly compared with all the templates in the appropriate age templates set. Each of the templates is reddened by Av=-5+Av$_t$,5+Av$_t$ mag with a step of 0.2 mag, where Av$_t$ is the extinction value from the literature. We use the extinction law from \citet{cardelli89} with R$_V$=3.1 to model the effect of extinction. In Appendix~\ref{appendix_ext_law} we show that the results are not affected greatly by a change in the extinction law used. The target spectrum is resampled to the spectral template wavelength grid. The comparison is performed simultaneously for the JHK bands if available (not all spectra have the three bands, in those cases the comparison is made on the available bands), where the bands are defined as: J=1.15-1.3, H=1.5-1.8, K=2-2.3 $\mu m$.

In order to minimize the effect of systematics from the heterogeneous nature of our dataset, we add to each comparison a grid of wavelengths at which the spectra is normalized, meaning that in total there are three fitted parameters: spectral type, extinction and normalization wavelength. For each object we minimize the $\chi^2$ parameter:
\begin{equation}
    \chi^2=\frac{1}{N-m} \sum_{i=1}^{N} (O_i-T_i)^2
\end{equation}

where O is the object spectrum, T the template spectrum, N the number of data points, and m the number of fitted parameters (m=3). For each point in the distribution of $\chi^2$ values, we select the wavelength normalization value as the one that minimizes the $\chi^2$, leaving two free parameters: spectral type and extinction.

\begin{figure*}[hbt!]
    \centering
    \includegraphics[width=\textwidth]{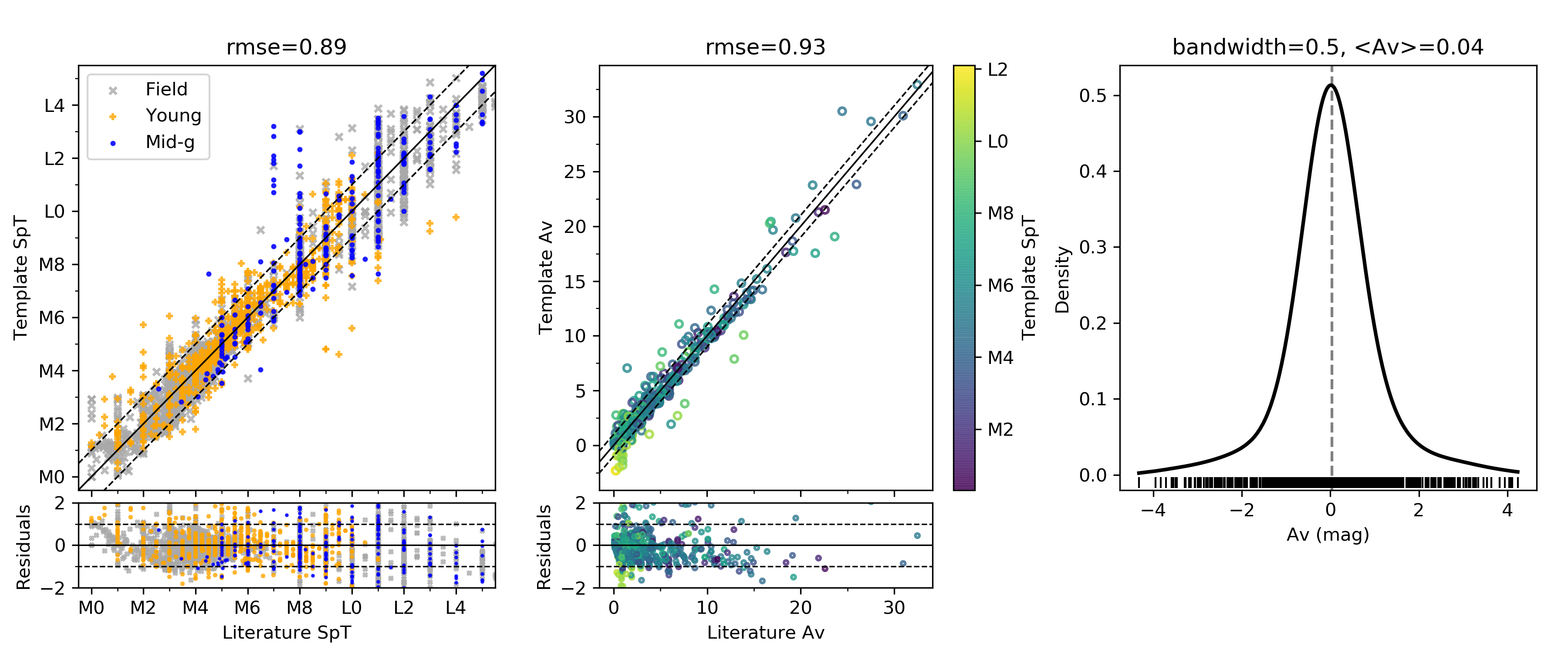}
    \caption{Results of the template fitting process explained in Section~\ref{templates}. Left panel shows the comparison of the literature SpT with the derived SpT and the residuals, grey crosses, blue circles and orange with the plus marker refer to the field, mid-gravity and young classes. Middle panel shows the extinction derived in this method compared with the literature value and the residuals, only for objects in star forming regions, colorcoded with the literature SpT. The right panel shows the KDE of the extinction for the low-extinction objects (field and NYMGs) with a bandwidth of 0.5. The mean value is shown in grey.}
    \label{fig:temp_results}
\end{figure*}

We convert the distribution of $\chi^2$ values into normalized probabilities, by applying the \textit{softmax} function \citep{goodfellow16} to 1/$\chi^2$/200 when using the young templates and to 1/$\chi^2$/600 when using the field templates (lower amount of templates). The \textit{softmax} function  transforms the given $\chi^2$ distribution into a normalized probability distribution proportional to the exponential of the input numbers. Finally, we derive the spectral type and extinction as the weighted average of the normalized probabilities of the best 20 fits. The uncertainty of these values is derived via error propagation: 

    \begin{equation}
        \sigma=\frac{1}{N}\sqrt{\frac{\sum_{i=1}^{N} w_i(x_i-Y)^2}{\sum w_i}}
    \end{equation}

where $N$ is the number of data points (20 in this case), $w$ are the normalized probabilities, $x$ are the associated values of the parameter that is being derived, and $Y$ is derived parameter.

Negative values are included in the extinction grid in order to avoid edge effects at A$_V$=0. Although field objects have negligible extinction, we decided to fit the extinction as well to minimize the effect of the heterogeneity of the dataset and therefore draw more reliable conclusions.

In Figure~\ref{fig:probmap} we show an example of the distribution of normalized probabilities for the case of a young object. The first thing we can notice is the known degeneracy between SpT and extinction. It also shows the SpT and extinction values derived in this section (in magenta), together with those from the literature (green) and the value derived in section~\ref{spt_indices} using spectral indices (cyan).

The left panel of Figure~\ref{fig:temp_results} shows the comparison between the literature and the derived SpT for the entire dataset. The three gravity classes are represented with different colors and markers. The two derivations are in good agreement, and the root-mean squared error (rmse) of this comparison is approximately 1. 
In the M0-M2 range the spectra are relatively featureless, due to the fact that the water bands that characterize the broadband shape of cool dwarfs, are not as prominent at these SpTs. This aspect makes spectral typing less accurate, which is what we observe in Figure~\ref{fig:temp_results}. The middle panel of Figure~\ref{fig:temp_results} shows the extinction from the literature compared with the value derived here for SFR members (non-negligible extinction). The rmse of the comparison is close to 1, mainly affected by highly extincted objects, where extinction derivation is more uncertain. We also observe that the extinction seems to be slightly underestimated when compared to the literature values from Av=5 mag. This effect may be associated with the conversion from the literature extinction which was measured in a different band (J or Ks) to Av. This discrepancy is of the order of 0.5 mag, which is the typical accuracy in the measurement of extinction. The right panel of Figure~\ref{fig:temp_results} shows the kernel density estimator of the extinction with a gaussian kernel of width 0.5, for the objects whose comparison was made with the field templates. The distribution is close to gaussian with a mean of $<$Av$>$=0.04 mag.

From here on we will focus on the objects with a derived SpT$<$L3, where we can better appreciate differences in the trends of the different gravity classes (there is no young object with SpT$>$L3 in the dataset), and we set the extinction to zero for those objects that had a derived negative extinction.

%--------------------------------------------------------------------
\subsection{Spectral type indices} \label{spt_indices}
%--------------------------------------------------------------------

In this section we evaluate the performance of various spectral indices defined for spectral typing in the literature, and define two new indices. Lastly we propose a method for SpT derivation using the selected spectral indices. The spectra have been corrected for the extinction estimated in the previous section using Cardelli's law with R$_V$=3.1.

We inspected the following indices defined in the literature: WH, WK, QH, QK \citep{weights09}, H2O \citep{allers07}, H2O-A, H2O-B, H2O-C, H2O-D, J-FeH \citep{mclean03}, H2O-1, H2O-2, FeH \citep{slesnick04}, HPI \citep{scholz12a}, sHJ, sKJ, sH2O$^{J}$
, sH2O$^{H1}$, sH2O$^{H2}$, sH2O$^{K}$ \citep{testi01}, H2O-K2 \citep{rojas-ayala12}, H2O-H, H2O-K \citep{covey10}, Q \citep{wilking99,cushing00}, H$_2$O (1.2 microns) \citep{geballe02}, wO, wD, w2 \citep{zhang18}. The indices and their functional forms are listed in Table~\ref{tab:indices_app} of the Appendix.

We evaluate the performance of all the indices on our dataset by comparing them with the SpT derived from spectral templates (section~\ref{templates}) and with the literature SpT.

Our requirements for a useful SpT spectral index are the following:
\begin{itemize}
    \item Good correlation with SpT, with a maximum error $\pm$1 sub-SpT. 
    \item Gravity-insensitive.
    \item Have a good performance in a SpT range of at least 5 subtypes within our range of interest (M0-L3).
\end{itemize}

\begin{figure*}[hbt!]
    \centering
    \includegraphics[width=\textwidth]{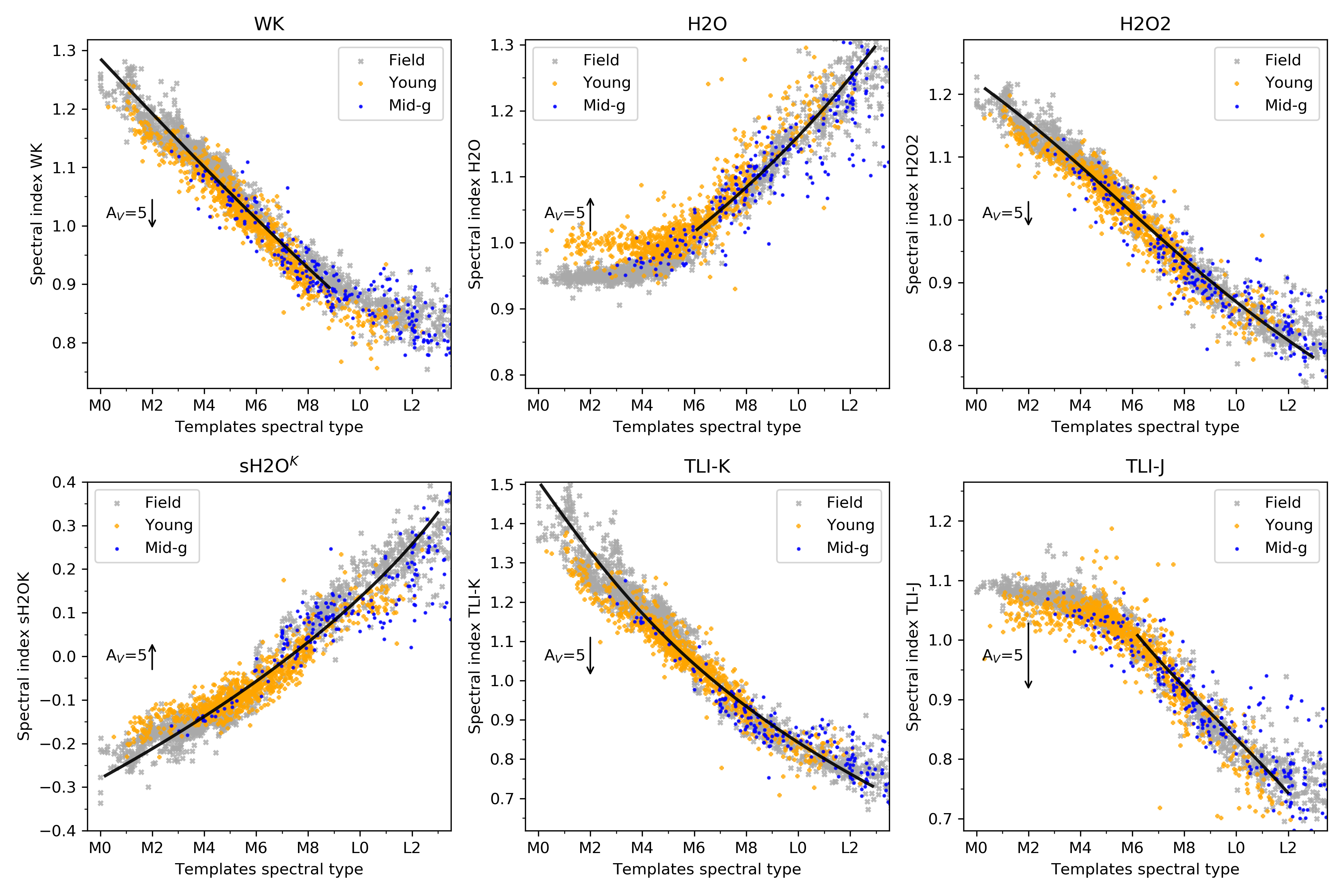}
    \caption{The selected literature and new spectral type indices, versus the spectral type derived from the comparison to templates (Section~\ref{templates}). Young, mid-gravity and field objects are shown in orange, blue and grey respectively. The derived calibration curve is shown as a black solid line, over the sensitivity range of each index. The black arrows show the effect that an addition of five magnitudes of extinction would have on the index values.}
    \label{fig:sptind}
\end{figure*}

Only four indices meet the requirements: H2O, H2O-2, WK and sH2O$^{K}$ (see Figure~\ref{fig:sptind}). Figure~\ref{fig:sptind} also shows the calibration curve in the range of applicability of the index. For the WK, H2O-2 and sH2O$^{K}$ indices we recalibrate their correlation with SpT using a third degree polynomial. The remaining SpT indices inspected are shown in Appendix~\ref{appendix_sptindices}. Table~\ref{tab:tab_ind_spt} shows the sensitivity range, the coefficients of the SpT calibration and the rmse of each selected index.

We then performed a systematic search for a new SpT-sensitive index in the entire range of the JHK bands with the following formulation:
\begin{equation}
    index=\frac{\langle F_{\lambda_1-\lambda_2} \rangle}{\langle F_{\lambda_3-\lambda_4} \rangle} 
\end{equation}

where the $\lambda_i$ identify the ranges of wavelengths over which the fluxes are computed.

We define two new spectral indices (see Table~\ref{tab:tab_ind_grav}): the first one is based on the flux ratio of two bands centered at 1.98 and 2.23 $\mu m$ with a bandwidth of 0.02 $\mu$m, we name it TLI\footnote{TLI stands for The Lost Index.}-K. The first band overlaps with the last two lines of the Calcium triplet at $\sim$1.97$\mu m$. The bottom middle panel of Figure~\ref{fig:sptind} shows its behaviour with SpT which is almost linear over the entire SpT range and shows a smaller dispersion than the other selected indices. The second index, named TLI-J, is based on the flux ratio of two bands centered at 1.205 and 1.28 $\mu m$ with a bandwidth of 0.01 and 0.02 $\mu m$ respectively. The first band overlaps with the FeH absorption band at $\sim$1.2 $\mu m$, a known gravity-sensitive feature (see section~\ref{youth}), however, the index does not show a gravity-sensitive behaviour. The bottom right panel shows its behaviour with SpT, which is linear from $\sim$M6 and presents a very low dispersion as well up to $\sim$L2 (see Table~\ref{tab:tab_ind_spt}). 
All the selected spectral indices are extinction-sensitive, as they are only based on two different spectral narrow bands, the effect of A$_V$=5 mag is shown in Figure~\ref{fig:sptind}.

\begin{table*}
    \caption{Spectral type indices.}
    \begin{center}
        \begin{tabular}{l c c c c c c c}
        \hline\hline
        \multirow{2}{*}{Index name} & Index & Sensitivity & \multicolumn{4}{c}{Coefficients\tablefootmark{a}} & rmse \\ \cline{4-7}
         & reference & range & c$_0$ & c$_1$ & c$_2$ & c$_3$ & (SpT) \\
        \hline
           H2O & \citet{allers07} & M6-L3 & -83.5437 & 169.388 & -104.424 & 24.0476 & 0.95 \\
           H2O-2 & \citet{slesnick04} & M0-L3 & 94.654 & -207.637 & 177.017 & -57.738 & 0.7 \\
           WK & \citet{lucas06} & M0-M9 & 39.944 & -48.659 & 19.597 & -4.596 & 0.57 \\ 
           sH2O$^{K}$ & \citet{testi01} & M0-L3 & 7.302 & 21.883 & -14.434 & 1.469 & 0.85 \\ 
           TLI-K & This work & M0-L3 & 48.82 & -75.354 & 42.669 & -9.396 & 0.68 \\ 
           TLI-J & This work & M6-L2 & -13.952 & 127.688 & -175.33 & 67.941 & 0.79 \\ 
        
        \hline
        \end{tabular}
    \tablefoot{$^{a}$ The spectral type is calculated as: SpT=c$_0$+c$_1 \cdot$(index) +c$_2 \cdot$(index)$^2$+c$_3\cdot$(index)$^3$}
    \end{center}
    \label{tab:tab_ind_spt}
\end{table*}

We propose a new method for spectral type derivation using the six selected spectral indices:
\begin{enumerate}
    \item An initial assessment of the SpT is done using the indices that perform well over the entire range of interest (M0-L3): WK, sH2O$^{K}$, H2O-2 and TLI-K. 
    
    \item Depending on the result, further refinement may be achieved:

\begin{itemize}
    \item SpT<M6: Keep the initial estimate.
    \item M6<SpT<M9: Use indices WK, H2O-2, sH2O$^{K}$, H2O, TLI-K, TLI-J.
    \item M9<SpT<L2: Use indices H2O-2, sH2O$^{K}$, H2O, TLI-K, TLI-J.
    \item L2<SpT<L3: Use indices H2O-2, sH2O$^{K}$, H2O, TLI-K.
\end{itemize}
\end{enumerate}

\begin{figure*}[hbt!]
    \centering
    \includegraphics[width=\textwidth]{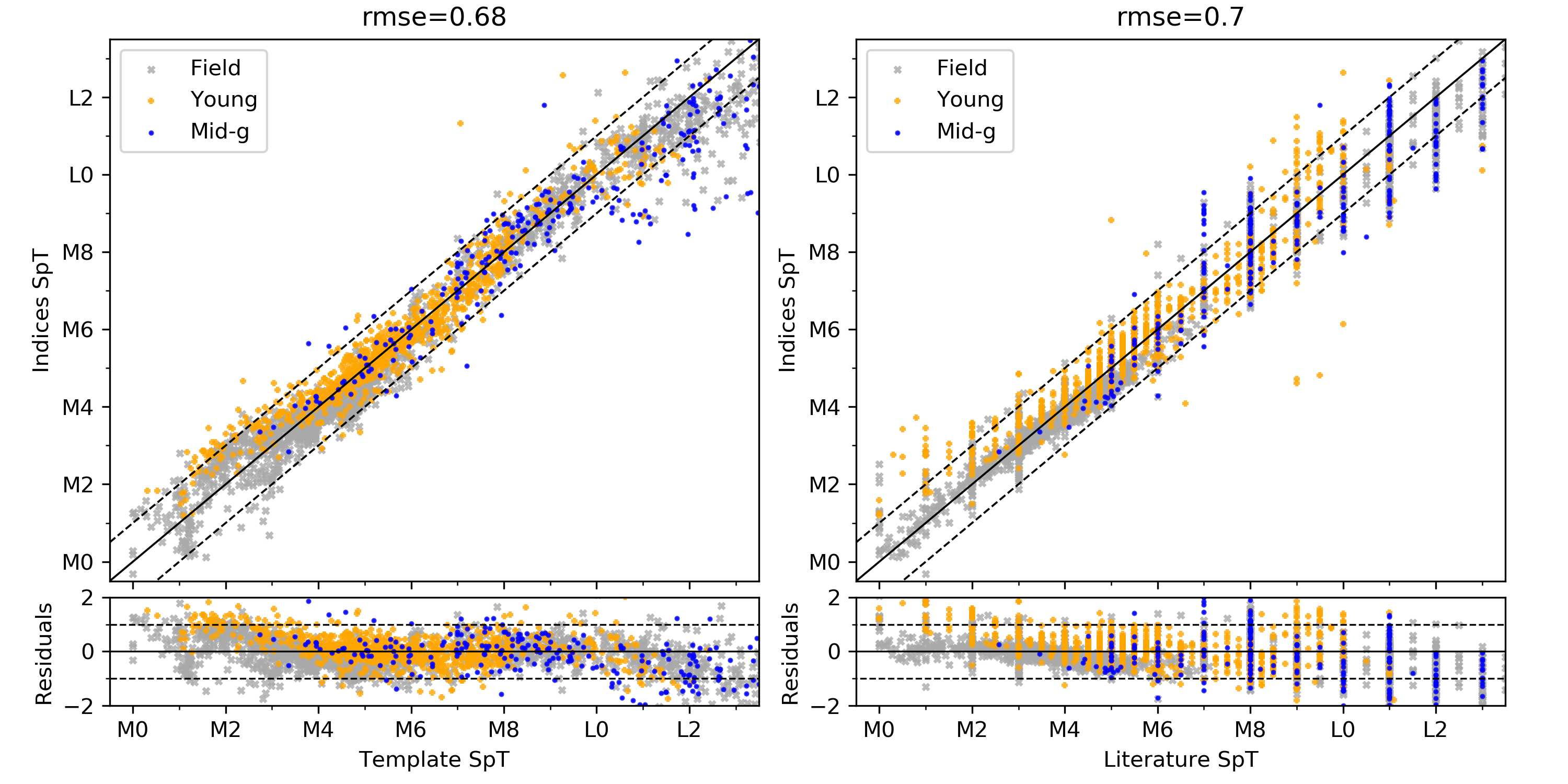}
    \caption{Comparison of the SpT derived using spectral indices (section~\ref{spt_indices}) with the SpT derived using spectral templates (section~\ref{templates}) and with the SpT from the literature. The black solid line represents the one-to-one correlation, and the dashed lines represent $\pm$1 subSpT. In the bottom region of the figure we show the residuals of the comparison.}
    \label{fig:spt_rec}
\end{figure*}

Figure~\ref{fig:spt_rec} shows the comparison of the SpT derived using this method with those derived using templates SpT (left panel, section~\ref{templates}) and the values from the literature (right panel). Both comparisons have a rmse significantly lower than 1, especially when compared with the values derived from templates, which we consider our reference. The range of calibration of this method is M0-L3, although in the M0-M2 range the flatness of the spectra makes the SpT derivation less certain, as in the template comparison case. The correlation of these indices with SpT starts to flatten at $\sim$L2 (see Figure~\ref{fig:sptind}). This aspect has an impact in the derivation of the SpT using these indices, in Figure~\ref{fig:spt_rec} we can see that from L2 the SpT is underestimated when compared with both the SpT derived by comparison with spectral templates, and from the literature.

The method employing spectral indices has the advantage of being mostly insensitive to gravity, unlike 
the template comparison where a correct age range needs to be taken into account.
Another advantage comes from cases where the spectrum has a limited wavelength range (one or two bands), which decreases the precision of the template fitting method,
and spectral indices may provide a more reliable measurement of the SpT.

\begin{figure*}[hbt!]
    \centering
    \includegraphics[width=\textwidth]{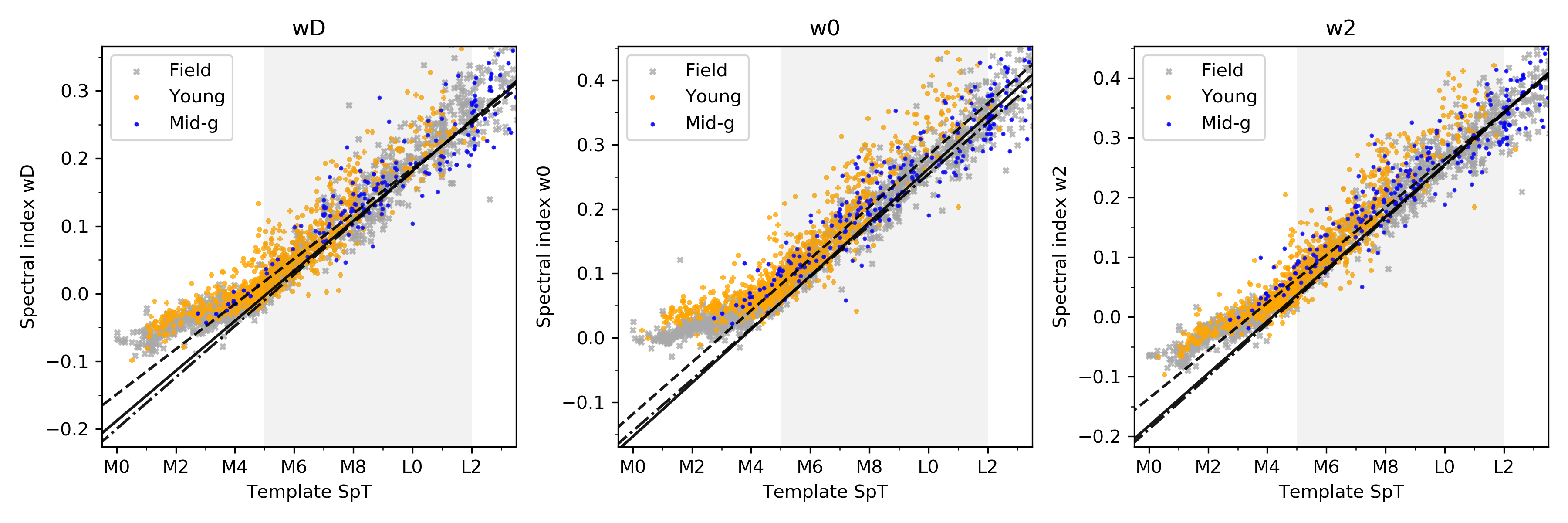}
    \includegraphics[width=\textwidth]{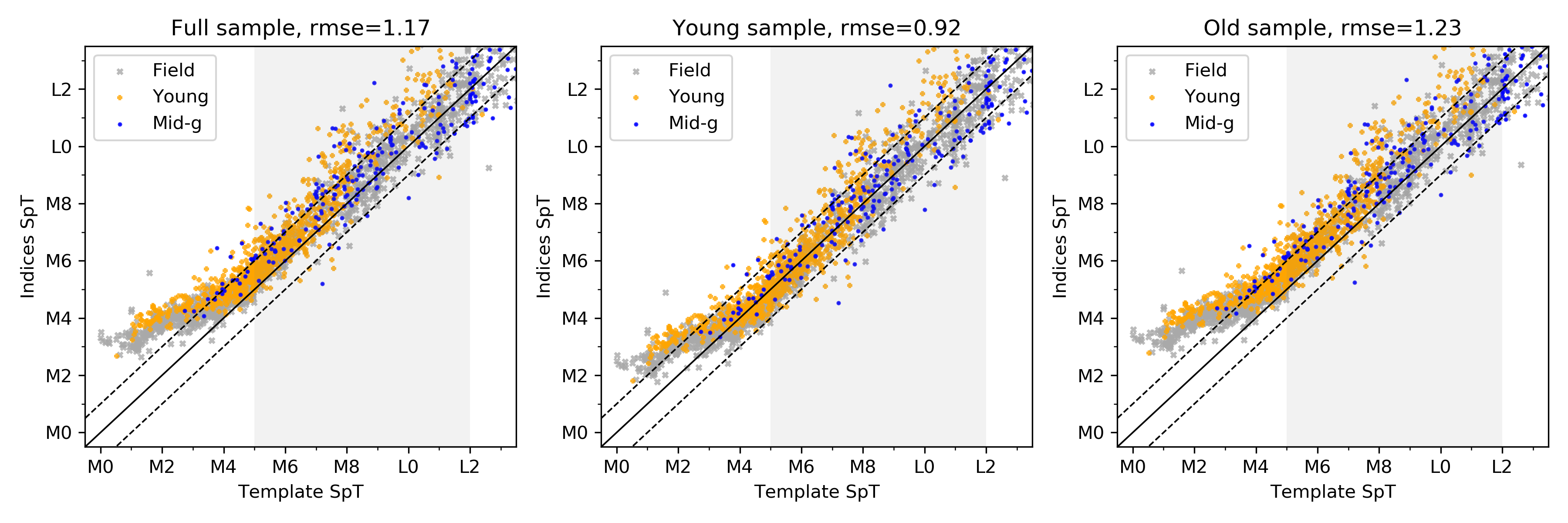}
    \caption{The top panels show the calibration of the \citet{zhang18} reddening-free SpT indices with the spectral type derived in Section~\ref{templates}. Young, mid-gravity and field objects are shown in orange, blue and grey respectively. The three available calibration curves are shown as solid (full sample), dashed (young sample) and dash dotted (field sample) black lines. The shaded grey region is the original range of application. The lower panels show the SpT derived for each index versus the SpT derived from the templates. The solid black line shows 1:1 linear relationship, while the dashed lines mark a difference of 1 SpT.}
    \label{fig:zhang_indices}
\end{figure*}

A disadvantage of using spectral indices for spectral typing is their sensitivity to extinction, which needs to be assessed with a separate method (e. g. template fitting). However, using more than two spectral bands and an extinction law, reddening-free SpT indices can be defined. This approach was originally devised in photometric studies via the Q index \citep{johnson53}, which has been later applied to the search of young low-mass stars and brown dwarfs \citep{najita00,allers20}. Various of the inspected spectral indices have the Q-like index formulation, such as QH, QK \citep{lucas06}, Q \citep{wilking99,cushing00}, and wO, wD and w2 \citep{zhang18}. We further analyse the performance of these indices, since they provide a very useful way of measuring the SpT for young extincted sources. The reddening-free spectral indices with the best behavior are the indices from \citet{zhang18}. \citet{zhang18} converted the four SpT indices selected in \allers~into three reddening-free SpT indices. The indices were defined for the M5-L2 SpT range. The top panels of Figure~\ref{fig:zhang_indices} show the indices together with the three calibration curves defined in \citet{zhang18} for the entire dataset in their range of application (M5-L2): full (solid), young (dashed) and field (dot dashed) sample. In the lower panels of Figure~\ref{fig:zhang_indices} we show the mean SpT derived using the three different calibration curves compared with the SpT derived in Section~\ref{templates}. The young sample fit is the one that produces the best results, in agreement with the original results. Using the young fit, these indices correlate well with SpT starting at M4, but from M8 young objects consistently appear over the +1 subSpT dashed line.

%--------------------------------------------------------------------
\section{Youth assessment with spectral indices} \label{youth}
%--------------------------------------------------------------------

In this section we evaluate the gravity-sensitive behaviour of cool dwarf spectra through spectral indices. In our wavelength range of interest (JHK bands) there are several gravity-sensitive features. In the J-band there is a NaI doublet at 1.14 $\mu m$, there are two KI doublets at 1.18 and 1.24 $\mu m$, and two FeH absorption bands at 1.2 and 1.24 $\mu m$. These features are more pronounced in the spectra of field objects. The H-band has a broadband gravity-sensitive behaviour where younger objects have a more triangular and sharp shape than field objects \citep{lucas01}. This feature has been explained by a decrease in H$_2$ collision induced absorption (CIA) with pressure \citep{kirkpatrick06}. In the H-band, we also find multiple FeH bands. In the K-band, gravity sensitive features are the NaI doublet at 2.21 $\mu m$, although its correlation with surface gravity is weak \citep{gorlova03}, as well as the broad shape of the spectrum, which peaks at different wavelengths for young and field objects \citep{canty13}. 

The main goal of the present work is to provide a relatively straightforward method which would allow to distinguish the young members of SFRs from other contaminants. The main source of contamination in these samples are reddened background sources with featureless spectra, that can be easily separated from cluster members. But in these samples there are also foreground low-mass stars, reddened background giants and even young field cool dwarfs with no known membership to any moving group. We therefore split our main goal in two:

\begin{itemize}
    \item Distinguish field and young cool dwarfs in the whole SpT range of interest (M0-L3).
    \item Test how well can the spectra of young SFR members be differentiated from somewhat older members of NYMGs.
\end{itemize}

\subsection{Gravity-sensitive spectral indices}

There are several gravity-sensitive spectral indices defined in the literature. We start by re-assessing their performance over our SpT range of interest. The gravity-sensitive index H$_2$(K) \citep{canty13} was defined using objects with SpT between M8 and L0 to model the change in the K-band peak from younger objects at $\sim$2.24 $\mu m$ to older objects at $\sim$2.17 $\mu m$. The HPI index \citep{scholz12a} was defined as a SpT-sensitive index for young objects, but presents a gravity-sensitive behaviour as well. This index uses a narrow band at the brightness peak of cool dwarfs at 1.68 $\mu m$. Lastly, \allers~ defined three indices that fall in our wavelength range of interest: the H-cont index based on the shape of the H-band, with three different narrow bands located centered at 1.47, 1.56 and 1.67 $\mu m$. The KI$_J$ index based on the KI doublet at 1.244 and 1.253 $\mu m$. And the FeH$_J$ based on the FeH absorption feature at 1.2 $\mu m$, which was defined for moderate resolution spectra (R$>$750).

\begin{figure*}[hbt!]
    \centering
    \includegraphics[width=\textwidth]{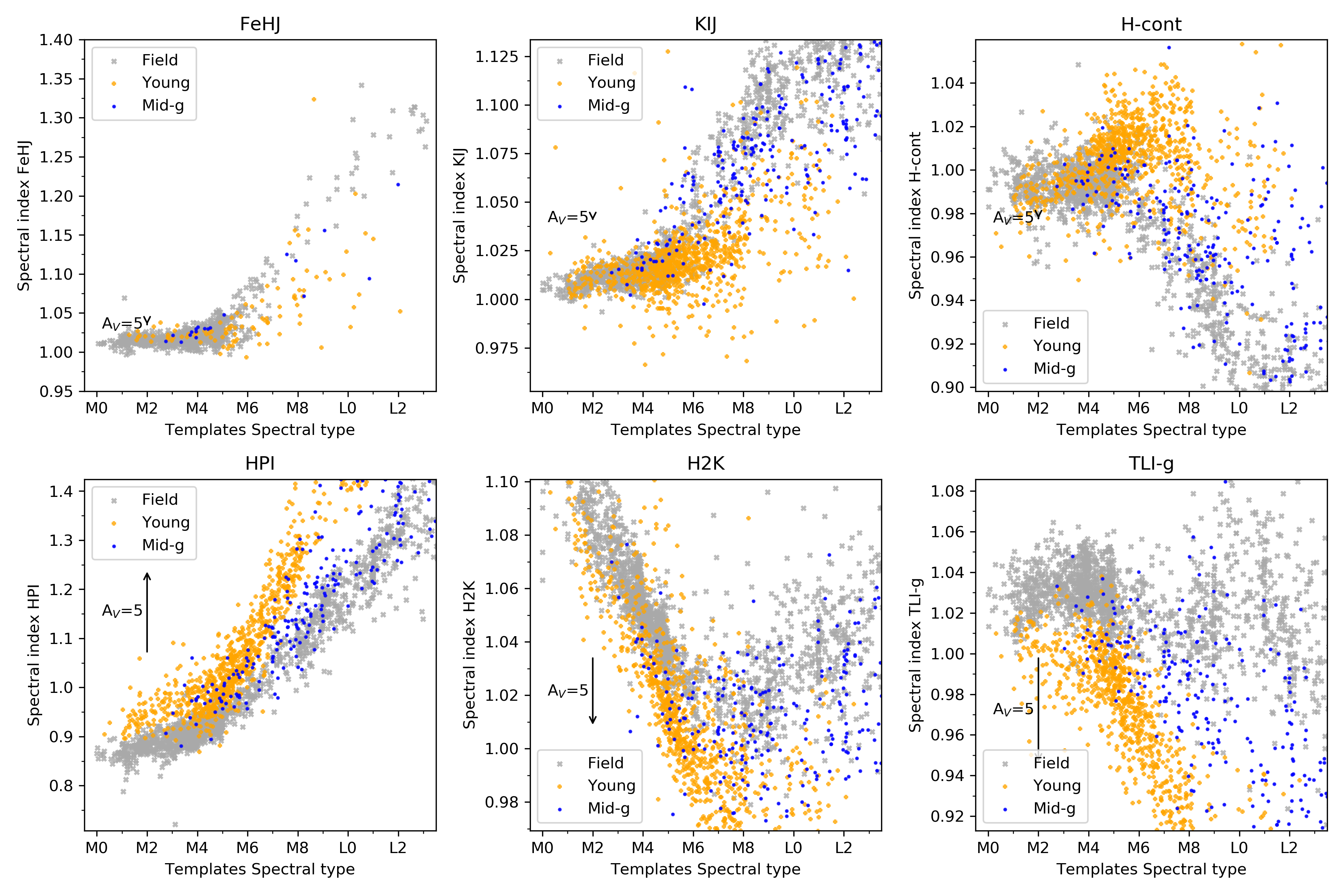}
    \caption{Gravity sensitive indices defined in the literature and the new index defined in this work compared with the SpT derived in section~\ref{templates}.}
    \label{fig:youthinds}
\end{figure*}

In order to test the behaviour of these indices, we use the homogeneously derived SpT and extinction from section~\ref{templates}. We divide the spectral library in the same three classes introduced in section~\ref{dataset}. Figure~\ref{fig:youthinds} shows the behaviour of the different indices from the literature compared with the SpT. The three age/gravity classes are represented by different colors and markers. The H-cont and KI$_J$ indices have similar behaviour, the young and field trends are indistinguishable until M6 where some separation starts to appear. The intermediate gravity objects overlap with both the young and field objects. The H$_2$(K) index has a similar behaviour as it was defined only for objects with SpT in M8-L0. The FeH$_J$ is only calculated for the moderate resolution spectra (R$>$750), no clear disentanglement of the age classes is observed. Lastly, the HPI index has a good correlation with SpT, and that makes it easier to separate the age classes, although we can see that this separation only starts at M6 and is not better than for the other indices.

After assessing the gravity-sensitive indices from the literature, we performed a systematic search for new gravity-sensitive indices along the whole wavelength range. We define a new gravity-sensitive index based on the 1.5-1.565 micron H$_2$O absorption band, which we name TLI-g, where g stands for gravity-sensitive (see Table~\ref{tab:tab_ind_grav}). As can be seen in the bottom right panel of Figure~\ref{fig:youthinds}, the separation between the young and field sequences is more pronounced than for any of the previously defined indices, and the mid-gravity objects clearly populate the intermediate space between both trends. The mid-gravity class contains NYMGs members, and field dwarfs with youth signatures, whose ages are difficult to determine. The field and young separation is clear from M2 at least (there are very few objects at earlier SpTs), and is maintained down to L2 (latest SpT of a young object in the dataset).

\begin{table}
    \caption{New spectral type and gravity spectral indices defined.}
	\begin{center}
        \begin{tabular}{l c c c c}       
            \hline\hline
            Index name & $\lambda_1$ & $\lambda_2$ & $\lambda_3$ & $\lambda_4$ \\
            \hline
            TLI-J & 1.20 & 1.21 & 1.27 & 1.29 \\
            TLI-K & 1.97 & 1.99 & 2.22 & 2.24 \\
            TLI-g & 1.56 & 1.58 & 1.625 & 1.635 \\
            \hline 
		\end{tabular}
	\end{center}
    \label{tab:tab_ind_grav}
\end{table}

\begin{figure*}[hbt!]
    \centering
    \includegraphics[width=\textwidth]{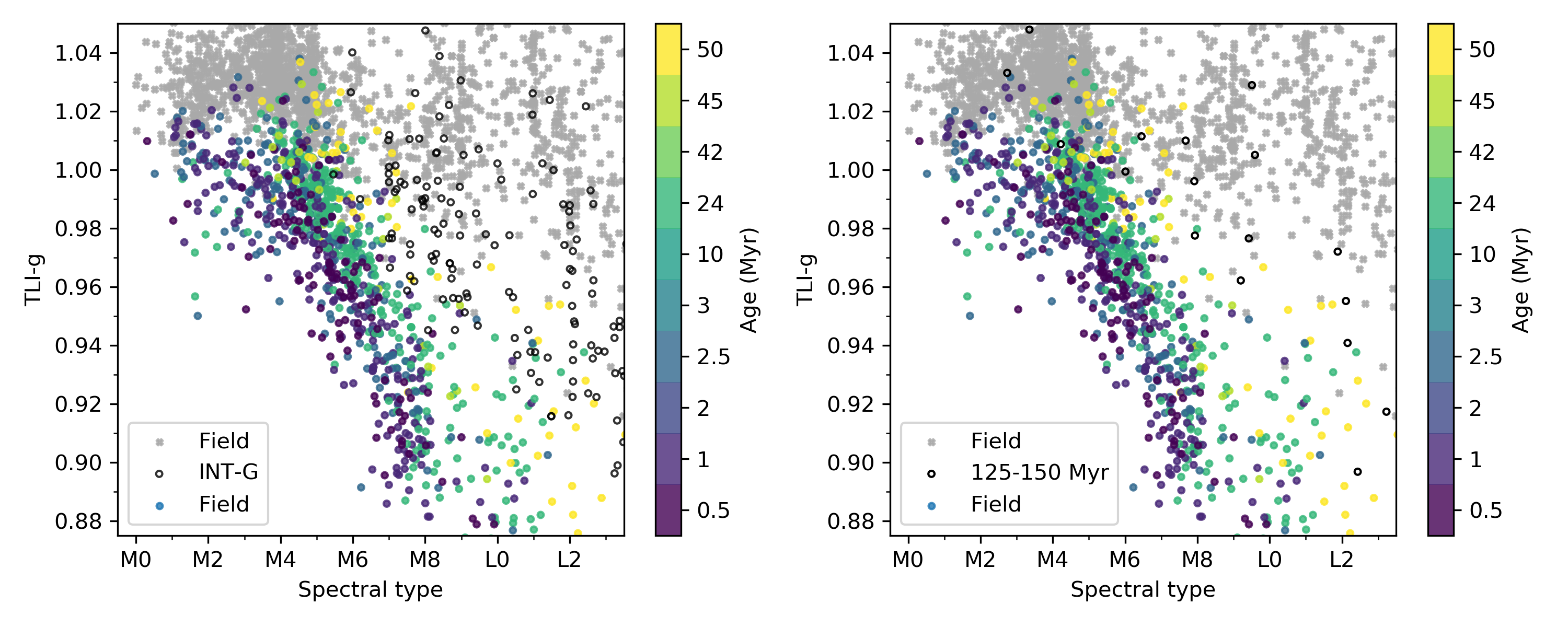}
    \caption{TLI-g index versus the SpT derived in section~\ref{templates}. In both panels we show field objects as grey crosses and young objects with known age are color-coded according to it. In the left panel we show INT-G objects from \allers~ without age determination as black circles. In the right panel we show objects from Pleiades and AB Dor (ages between 125 and 150 Myr) as black circles.}
    \label{fig:tliage}
\end{figure*}

The trend of the different age classes for the TLI-g index seems to maintain at the late SpTs. This is very interesting as growing numbers of L-type SFR members are now being found \citep{bayo11,lodieu18} and there are even some young T-type candidates proposed \citep{pena15}. Being able to disentangle field and young populations at mid to late-L SpTs using low-resolution spectra will prove of the utmost importance in the search of the lowest-mass substellar objects in SFRs, whose populations will be revealed shortly thanks to the JWST. 

We also inspect the age calibration of the TLI-g index. In both panels of Figure~\ref{fig:tliage} we show field objects as grey crosses and objects younger than 50 Myr are color-coded according to their respective ages. As black open circles we show in the left panel objects with INT-G gravity classification (\allers) without known age, and in the right panel objects with ages between 125 and 150 Myr (members of Pleiades and AB Dor NYMG). Objects with ages below 25 Myr cluster close together, and as the age increases the objects get further away from the young sequence, and closer to the field population. Objects with ages of 125 and 150 Myr (from Pleaides and AB Dor respectively) overlap fully with the field sequence. A large fraction of the INT-G objects overlap with the field trend, but there is also a significant amount of them are located close to the 50 Myr objects.

High metallicity can mimic some of the low-gravity spectroscopic features \citep{shkolnik09}, for example, the H-band has been seen to have an enhanced low gravity shape in some dusty brown dwarfs \citep{looper08}. At the same time, these objects would show a field-like strength in other youth features, such as alkali lines. This behaviour is attributed to a high atmospheric dust content or thicker clouds, since the triangular H-band shape is likely caused by a reduced collision induced hydrogen absorption and/or an increased condensate opacity \citep{borysow97,kirkpatrick06,looper08}. Low gravity objects can also have a high dust content, so a clear distinction between the two scenarios may not always be possible \citep{kellogg15}. Because of this, \allers~caution against using the H-band as the only gravity feature, as they claim that their H-cont index classifies these dusty objects as low gravity. However, in Figure 24 of \allers~dusty objects lie very close or inside the field gravity boundary for the mentioned H-cont index. These dusty objects typically have late-M to late-L SpTs, present peculiar red colors, as the J-band is typically suppressed compared with the HK bands, and they do not fit well low-gravity, field-like nor low metallicity templates \citep{looper08,kirkpatrick10}.

Whenever possible, we encourage the usage of more than one youth feature to classify for youth/gravity, as even if the H-band can appear more pronounced for some peculiar objects, the rest of the gravity sensitive features should look field like. On the other hand, there are very few known objects that meet the dusty criteria, and not all the objects flagged as being dusty or red peculiar present this feature, so the possibility of finding a red dusty object in a sample of SFR candidates is negligible.

The performance of the TLI-g index is further evaluated in the next section. We have shown that the TLI-g index has a better age/gravity class separation than any index previously defined, and it is based on a broadband feature, allowing its calculation on very low resolution spectra. Using this index alone it is not possible to distinguish three gravity classes, but the youth trend is maintained up to at least 50 Myr, making it particularly suitable for decontamination of young cluster member sequences. It remains to study its behaviour at later SpTs, as there are very few to none young objects at SpTs later than L3, although it seems that the tendency may be maintained.

%--------------------------------------------------------------------
\section{Youth assessment with machine learning} \label{mlearning}
%--------------------------------------------------------------------

\begin{figure*}[hbt!]
    \centering
    \includegraphics[width=\textwidth]{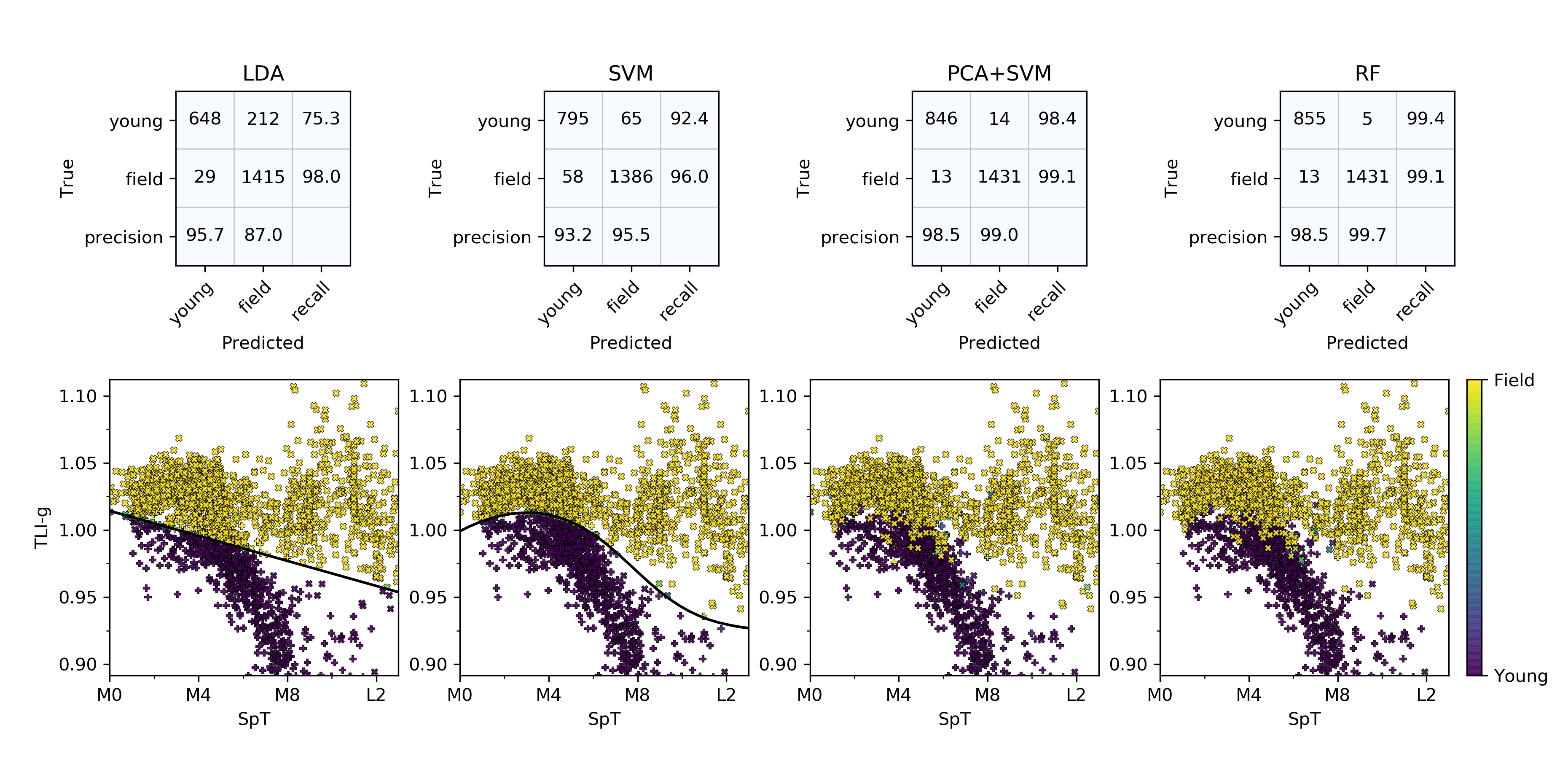}
    \caption{Machine learning results predicting for two age classes (young and field). The input for the LDA and SVM methods is the TLI-g index, while PCA+SVM and RF methods consider the entire JHK spectrum. The top panels show the confusion matrices and precision and recall values obtained. The bottom panels show the TLI-g versus SpT for the dataset, where the color coding is the age class probability. For the LDA and SVM models the decision function is shown as a solid black line.}
    \label{fig:cm_y}
\end{figure*}

Machine learning methods can also in principle be adopted for the discovery of new features and/or to directly perform youth inferences, but so far, stellar youth spectroscopic features have only been studied via spectral indices and visual inspection. Machine learning can provide an efficient approach to find relationships between a large number of variables, enabling supervised and unsupervised classification, and also to reduce the complexity of a given problem by discovering lower dimensional spaces that are more amenable to interpretation \citep[e.g.][]{2019arXiv190407248B}. These methods have been applied on photometry and astrometry to identify members of open clusters \citep[e.g.][]{krone14,castro18}, identify young stellar objects on wide sky areas \citep[e.g.][]{marton16,marton19, 2021ApJS..254...33K} or to identify new members of star forming regions \citep[e.g.][]{Ducourant2017,melton20,galli20}. They have also been applied directly to stellar spectroscopy for the derivation of  stellar parameters \citep[e.g][]{sarro18,olney20,passegger20}.

In this section we adopt machine learning methods to perform age/gravity classification from very low-resolution spectroscopy, and perhaps more importantly, to help the identification of new spectral features that correlate with stellar youth.
Here we use supervised learning methods. Thus, we start from a training set that, ideally, has a coverage of the parameter space \citep[e.g.][]{2017MNRAS.468.4323B} and, in the case of the methods adopted herein, to be in the same units/scale, with homogeneous sampling, and to contain no missing data. Thus the following homogenization steps are applied to the original data:
\begin{itemize}
    \item SpT and extinction are taken from Section~\ref{templates}. All spectra are de-reddened with this extinction value.
    \item We resample the dataset to the same wavelength grid. The spectral range consists of the JHK bands neglecting the telluric regions between them: $\lambda\in$[1.15-1.35,1.5-1.78,2-2.3] $\mu m$, with a wavelength step $\Delta \lambda = 0.043 \mu m$ (the largest in the dataset).
    \item All the objects are scaled to an integrated flux of one.
\end{itemize}

We also require good spectral quality over the entire wavelength range, reducing our original dataset to 2528 spectra.

We compare the performance of four different models: Linear Discriminant Analysis (LDA), Support Vector Machine (SVM), Principal Component Analysis (PCA)+SVM and Random Forests (RF). The first two models are applied to the previously computed TLI-g index and SpT, while we use the entire spectra as input for the two latter methods. We apply these models on two different case scenarios: distinguishing between two (young and field) and three (young, mid-gravity and field) age classes. We also test the performance of the PCA+SVM and RF models using only the HK bands portion of the spectrum.

The details about the models, the selection of the hyper-parameters and the procedure followed for the evaluation of the performance are given in Appendix~\ref{appendix_ml}.

%--------------------------------------------------------------------
\subsection{Two age classes}
%--------------------------------------------------------------------

\begin{figure*}[hbt!]
    \centering
    \includegraphics[width=\textwidth]{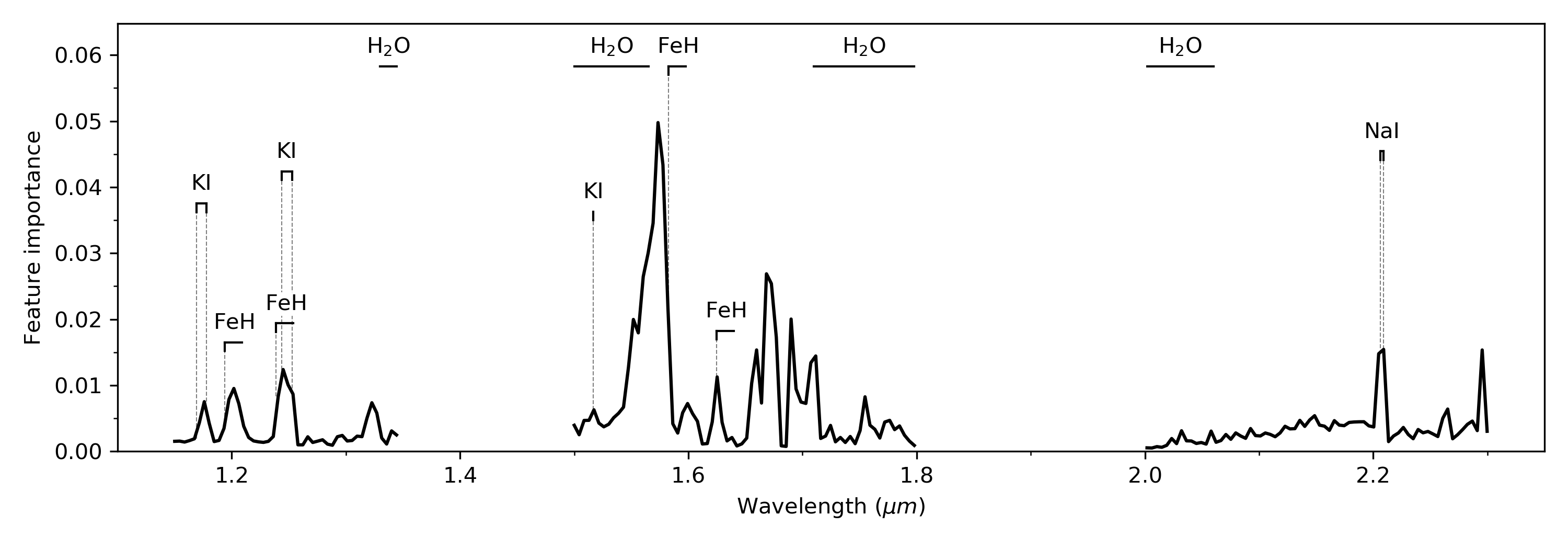}
    \caption{RF feature importance for the entire JHK range, when classifying for two age classes: field and young. We also show the known gravity-sensitive features as well as the water bands, the main driver of the cool dwarfs broadband shape.}
    \label{fig:importance_JHK_2}
\end{figure*}

\begin{figure}[hbt!]
    \centering
    \includegraphics[width=\textwidth/21*10]{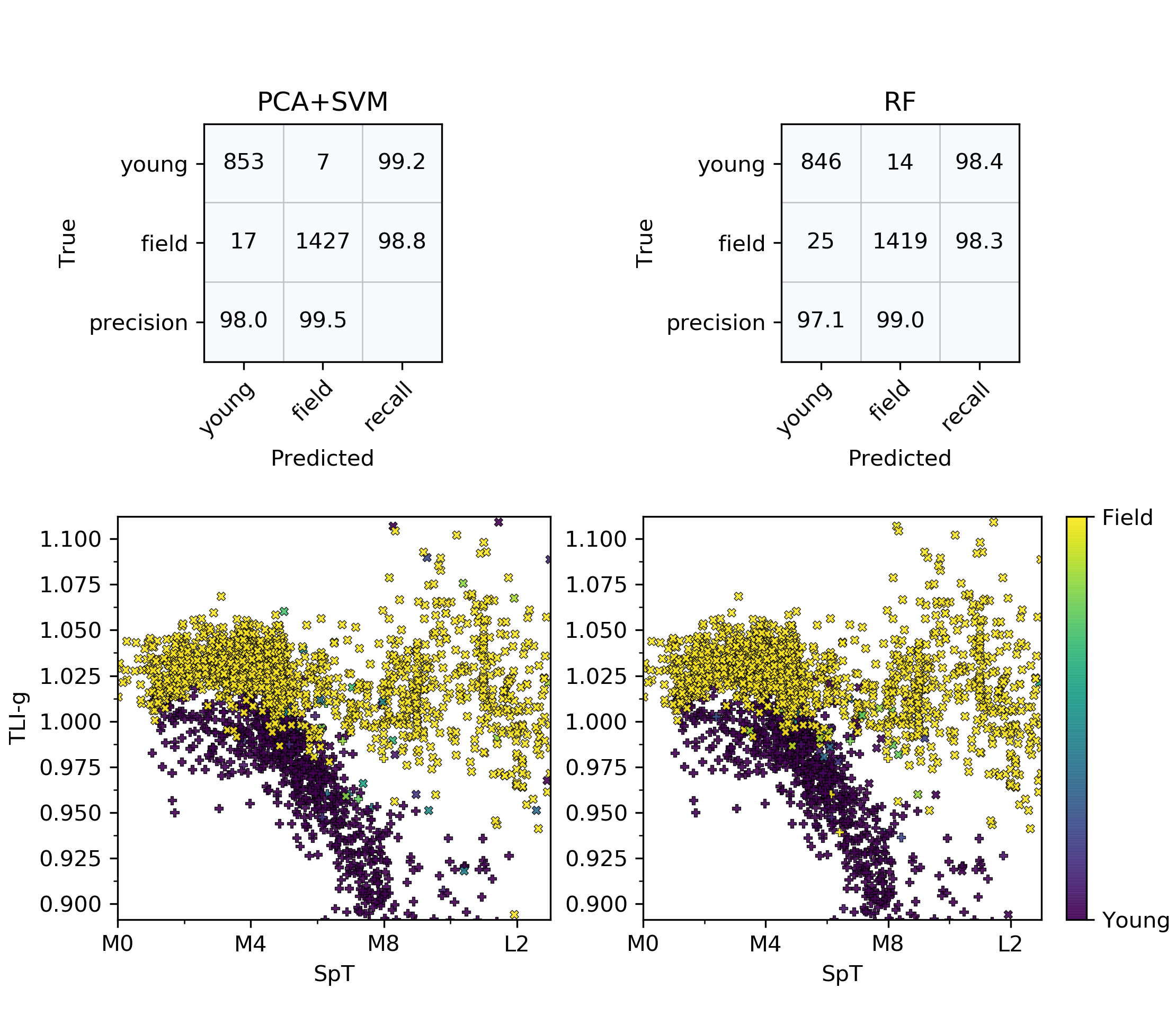}
    \caption{Machine learning results predicting for two age classes (young and field) using the HK bands. The top panels show the confusion matrices and precision and recall values obtained. The bottom panels show the TLI-g versus SpT for the dataset, where the color coding is the mean age class score.}
    \label{fig:cm_y_HK}
\end{figure}

\begin{figure*}[hbt!]
    \centering
    \includegraphics[width=\textwidth]{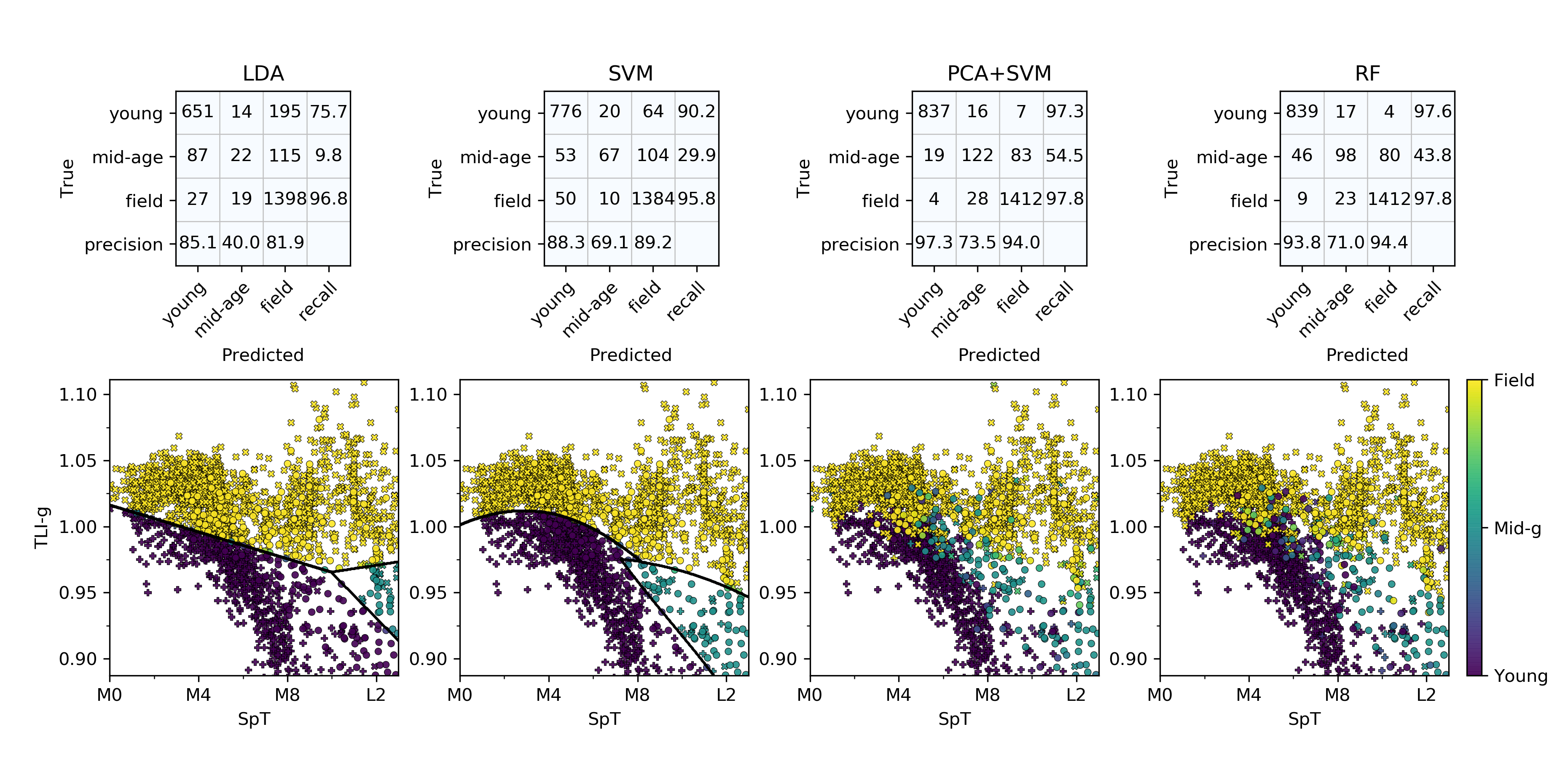}
    \caption{Machine learning results predicting for three age classes (young, mid-age and field) using the JHK bands. The top panels show the confusion matrices and precision and recall values obtained. The bottom panels show the TLI-g versus SpT for the dataset, where the color coding is the mean age class score. For the LDA and SVM models the decision function is shown in a solid black line.}
    \label{fig:cm_ymid}
\end{figure*}

\begin{figure}[hbt!]
    \centering
    \includegraphics[width=\textwidth/21*10]{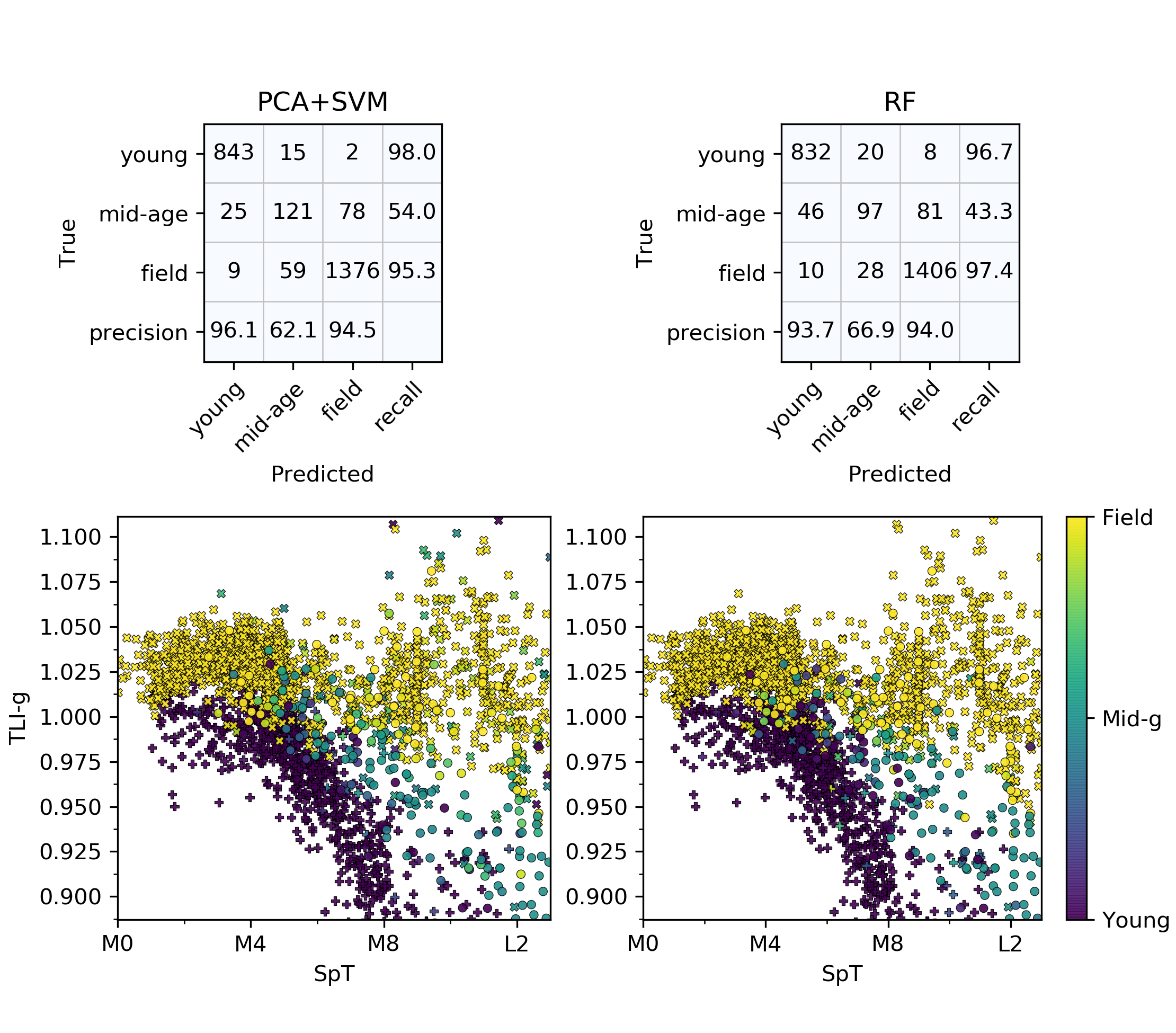}
    \caption{Machine learning results predicting for three age classes (young and field) using the HK bands. The top panels show the confusion matrices and precision and recall values obtained. The bottom panels show the TLI-g versus SpT for the dataset, where the color coding is the age class mean score.}
    \label{fig:cm_ymid_HK}
\end{figure}

We first applied the models predicting for two age classes: field (1444 spectra) and young (860  spectra). The classes are defined as scores (0 for young, and 2 for field), and the boundary for classification is made at a mean score of 1. The top panels in Figure~\ref{fig:cm_y} show the performance of the four models. The predictions are close to 100 \% for the SVM, PCA+SVM and RF models, but the two models operating on the entire spectra perform best. In the bottom panels of Figure~\ref{fig:cm_y} we present the TLI-g vs SpT diagram color-coded by the mean score of the 200 model runs, and also present the decision boundary of the LDA and SVM models for one run. LDA performs a simple linear decision curve, which does not work well for SpT$<$M5. The SVM method is able to separate both classes consistently in the entire SpT range, meaning that we can successfully recover almost all the young objects using only the TLI-g index and the SpT. RF and PCA+SVM results are similar to the SVM for most of the objects, but they perform better in regions of complete overlap of age classes in the TLI-g vs SpT space, having an overall separation of the classes that is almost perfect.

Using RFs we can estimate what are the most important features for the classification in age classes. This attribute is called feature importance. For each feature, RF derives the total reduction of the chosen criterion for tree split quality, assuming that the most important features produce the highest purity tree splits. If we evaluate this attribute with the RF model, we can study which parts of the spectrum are the most important when classifying for youth. In Figure~\ref{fig:importance_JHK_2} we show the feature importance for the two age classes case. The J-band has three big importance peaks at 1.18, 1.20 and 1.24 $\mu m$, the first and the third peaks overlap with KI doublets, and the second and third ones overlap with FeH bands. The three peaks are known gravity-sensitive features. Although these features are not fully resolved at these very low resolution spectra, they still affect its broadband shape. The H-band has a very prominent peak at 1.56-1.58 $\mu m$, and then three peaks at 1.63, 1.66-1.68 and 1.69-1.71 $\mu m$. The first two importance peaks lie very close to the center of the two bands used in the TLI-g index. The third and fourth peaks are close to the peak in flux for young cool dwarfs, and is included as one of the bands of the HPI index \citep{scholz12a}. The peaks at $\sim$1.57, 1.67 and 1.7 are associated with the broadband H-band gravity-sensitive behaviour, and the importance peak at $\sim$1.63 $\mu m$ coincides with the location of a FeH band. At 1.51 $\mu m$ there is a small peak that could be associated with a blended KI doublet. This peak maintains its relative importance compared with the most prominent peak (at $\sim$1.57 $\mu m$) in all cases inspected. The K-band has a uniform importance except for a peak at $\sim$2.2 $\mu m$ than can be associated with a NaI doublet. There is also a peak at the edge of the K-band which is most likely associated with an edge effect of the wavelength grid. Based on the importance map, we can also explore whether there are any obvious biases due to the heterogeneous nature of our dataset. These biases would appear as peaks in regions where no gravity-sensitive features are known to exist. For example, in a previous version of the dataset we observed unexpected importance peaks at the border of the bands, that were found to be associated with the rebinning process and not associated with youth features. The feature importance map helped us decide on the final wavelength range of the different bands.

Another test we perform to check for possible biases in the dataset, is to inspect the projection in the different principal components of the dataset and compared it with the SpT. A non-expected source of variance should appear separated from the general trend, and would be associated with a particular subset of the dataset. We find no such behaviour within the first 15 principal components.

These two tests show us that the heterogeneity of the dataset does not have an important effect when classifying for youth using the whole dataset. It means that the PCA+SVM and RF methods are actually identifying the youth features and using them for the classification.

We also test the results of the PCA+SVM and RF models using only the HK bands (see Figure~\ref{fig:cm_y_HK}), we obtain the same results within errors as in the JHK bands case. The feature importance plot (Figure~\ref{fig:importance_HK_2}) looks similar to the JHK case. The importance peak at $\sim$1.57 $\mu m$ is now even more important and the K-band NaI doublet importance is now blended with the flat importance of the entire K-band.

%--------------------------------------------------------------------
\subsection{Three age classes} \label{3ageclasses}
%--------------------------------------------------------------------

In this section, we evaluate the performance of the same four models as before, now classifying into three age classes. The classes are again defined as scores (0 for young, 1 for mid-gravity and 2 for field), and the boundaries for classification are made at a mean score of 0.5 and 1.5. We divide our dataset in the same three age classes of Section~\ref{youth}, having 860 young, 224 mid-gravity and 1444 field objects.

We follow the same procedure as for the two age classes case and show the results in Figure~\ref{fig:cm_ymid}. LDA and SVM are not able to disentangle the mid-age class for SpTs earlier than M6, and in that range the field/young division is very similar as for the two age classes case. From M6-M7 they are able to divide into three age classes, although the classes are not completely separable in the TLI-g vs SpT space. On the other hand, the results using PCA+SVM and RF are far better, we can see that all the objects that lie in between the field and young trends in the TLI-g vs SpT space are correctly classified as mid-gravity, while some of the objects lying over the young trend are also classified as mid-gravity. On the other hand, most of the mid-gravity objects lying over the field trend, are classified as field.

The importance of the features is similar as in the two age class scenario (Figure~\ref{fig:importance_JHK_3}). In the J-band, the most important peak is now the one located at $\sim$1.24 $\mu m$. There is another prominent peak at the edge of the J-band which could be associated with a difference in the shape of the H$_2$O absorption band with age, or with a wavelength grid edge effect like the one found at the end of the K-band. The peak at $\sim$1.63 $\mu m$ is much smaller and the one at $\sim$1.67 $\mu m$ is more prominent. The prominent peak at $\sim$1.57 $\mu m$ is divided in two, and is broader. The K-band maintains a mostly unifom importance and the NaI doublet peak.

We test the same classification using only the HK bands (see Figure~\ref{fig:cm_ymid_HK}). The results are again similar to the ones obtained with the JHK bands. The importance of the features (Figure~\ref{fig:importance_HK_3}) is mostly like the one obtained in the JHK bands case. This shows that if only HK bands are available, they can perform similarly as using JHK bands even at very low resolution. 

%--------------------------------------------------------------------
\section{Summary and conclusions}
\label{summary}
%--------------------------------------------------------------------

In this work we built a dataset that includes all the available near infrared spectra of cool dwarfs with SpT in the M0 - L3 range. The dataset includes 10 spectra of young brown dwarfs (SpT in M7-L2) obtained with SINFONI/VLT here presented.

We first inspected SpT derivation using spectral templates and spectral indices. By comparison with spectral templates we achieved a $\sim$1 subtype precision in SpT derivation, and a $\sim$1 mag precision in extinction, when compared with the values from the literature. We also inspected the behaviour of a large number of SpT-sensitve spectral indices from the literature, and defined two new indices (named TLI-J and TLI-K) that perform similarly to the best indices from the literature, in terms of the sensitivity range and the spread in the derived SpTs. Using six selected SpT indices, including the two defined in this work, we retrieved the SpT for the entire dataset with a precision below 1 subtype. The selected spectral indices are insensitive to variations in surface gravity but are sensitive to extinction. Their usage can be interesting when having a restricted spectral wavelength range.

The main motivation for this work was the evaluation of youth-sensitive features in low-resolution NIR spectra, that can be used to consistently separate the SFR members from field contaminants. To do so, we investigated previously proposed gravity-sensitive indices, and find that their utility is typically restricted to smaller SpT ranges than that explored in this work. We then defined a new gravity-sensitive index (TLI-g), which outperforms any other gravity-sensitive index previously defined. Using this index alone, the field and young ($\lesssim$ 10 Myr) classes are almost completely separable over the entire SpT range (M0-L3). Interestingly, the separation is progressively larger for the later SpTs,  making the TLI-g index potentially a very useful tool for the future surveys searching for the planetary-mass members of young clusters and SFRs, in case that the trend maintains for the SpTs later than L3. The mid-gravity class, composed of field objects with youth features and objects from NYMGs with ages above 10 Myr, populates the space in between the field and young classes in the TLI-g vs. SpT diagram, but also shows a slight overlap with them. The objects with INT-G classification from \allers~ gravity lie both over the field and mid-gravity trends. 
Objects with ages below 50 Myr cluster close to the young sequence and show a clear age-related gradient.

Finally, we also evaluated youth spectroscopic features using four machine learning models, testing how well we can separate two (young and field) or three (young, mid-gravity, and field) classes. To that end, we first employed LDA and SVM methods on the TLI-g vs. SpT plane, followed by PCA+SVM and RF, which use the entire JHK spectra as an input. All the four models successfully separate two classes (field and young), however, the models which use the full spectra have a slightly better performance than those using the TLI-g index, with metrics close to 100\%. When classifying for three age classes (field, mid-gravity and young), using the TLI-g index and the SpT alone, we are not able to effectively disentangle all the three classes. On the other hand, PCA+SVM and RF start to disentangle the mid-gravity class, although there is still a significant overlap with both the field and young sequences. Most of the mid-gravity objects classified as young are objects from NYMGs, and objects classified as field are mostly field objects with INT-G classification. 

We also tested the PCA+SVM and RF models on a wavelength range spanning only the H and K bands, which is commonly used in studies of cool dwarfs, and show that the results obtained in this case are similar to those using the full JHK range, for the predictions of both two and three age classes. Using RF feature importances we observe that the models are independently learning the most relevant features for youth determination. We find that the most important feature for youth classification is the H-band broadband shape, with the most prominent importance windows located at 1.56-1.58, 1.66-1.68 and 1.69-1.71 $\mu m$. Then, the FeH absorption bands at 1.2 $\mu m$ and 1.24 $\mu m$ and the KI doublet at 1.24 $\mu m$ are similarly important in all cases studied. The well-known gravity-sensitive feature caused by KI absorption at $\sim$1.18 $\mu m$ only appears important when classifying for two age classes. Further gravity-sensitive spectral regions are FeH band at 1.63 $\mu m$, the NaI doublet at $\sim$2.2 $\mu m$ and the $\sim$1.51 $\mu m$ feature associated with a blended doublet of KI, whose role in youth classification of cool dwarfs has not been discussed much in the literature. Interestingly, except for the NaI doublet, we find the K-band to have a flat importance. These results can prove very useful for: a deeper characterization of young clusters in extreme environments with upcoming multi-object spectrograph facilities (e.g. MOONS/VLT, NIRSPEC and NIRISS/JWST), the search and characterization of the low-mass end of the IMF with upcoming facilities (ELT and JWST) and also in the context of large scale NIR spectroscopic efforts, such as the ATLAS probe \citep{2019PASA...36...15W}.

Both the models applied on the entire spectrum (PCA+SVM and RF) and the models applied on the TLI-g index and the SpT (LDA and SVM) inspect broadband features that are driven by the variation of the surface gravity with age. The surface gravity of cool dwarfs goes from log($g$)$\sim$3.5 for the youngest objects to log($g$)$\sim$5 for field objects. The timescale of this change depends on the mass of the object.

Empirical measurements of the surface gravity using theoretical models show that in Pleiades ($\sim$120 Myr), the late-M objects already have a log$g$ value close to that of field objects and that they both get classified as INT-G and FLD-G \citep{martin17,manjavacas20}. It is therefore not surprising that the field dwarfs with youth signatures overlap with the old field dwarfs in both the TLI-g index and the machine learning models exploiting the full spectra.

\begin{acknowledgements}
   We thank the anonymous referee for her/his comments and suggestions. V.A-A., K.M. and K.K. acknowledge funding by the Science and Technology Foundation of Portugal (FCT), grants No. IF/00194/2015, PTDC/FIS-AST/28731/2017, UIDB/00099/2020 and SFRH/BD/143433/2019. A.K.M. additionally acknowledges the support from the FCT grant No. PTDC/FIS-AST/31546/2017.
\end{acknowledgements}

\bibliographystyle{aa} 
\bibliography{youth}

\begin{appendix}

\section{Planemos SINFONI} \label{appendix_sinfoni}

\begin{figure*}[hbt!]
    \centering
    \includegraphics[width=0.95\textwidth]{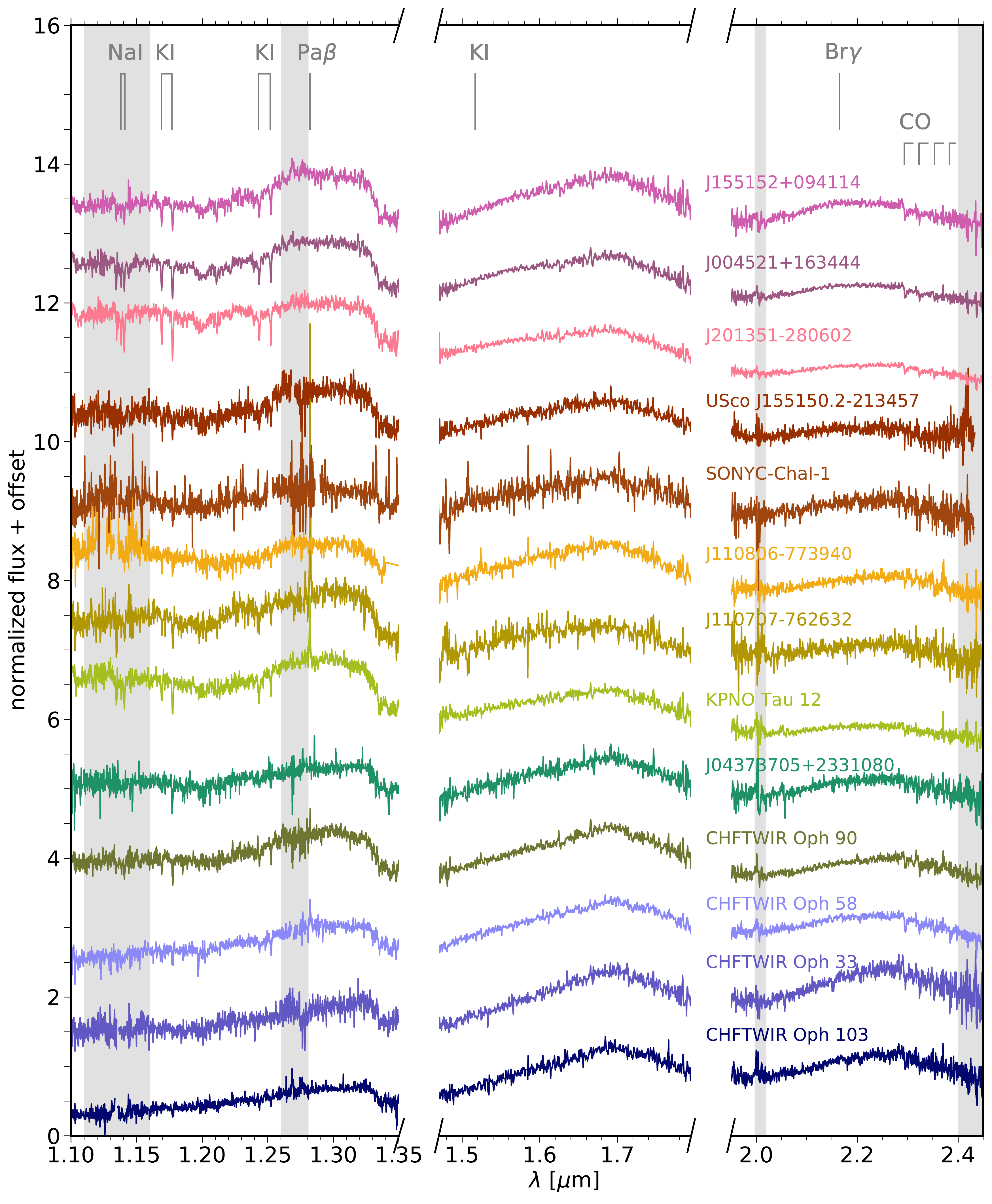}
    \caption{$JHK$ spectra of the objects observed with SINFONI/VLT. The top three spectra belong to field dwarfs, whereas the remaining are young members of star forming regions. The prominent lines in late-type spectra are marked. The grey shaded areas signal the regions with significant telluric absorption.}
    \label{fig:appendix_sinfoni}
\end{figure*}

In Figure~\ref{fig:appendix_sinfoni} we show the spectra observed with SINFONI presented in Section~\ref{sinfoni}.

\section{Extinction law}
\label{appendix_ext_law}

\begin{figure*}[hbt!]
    \centering
    \includegraphics[page=1, width=\textwidth]{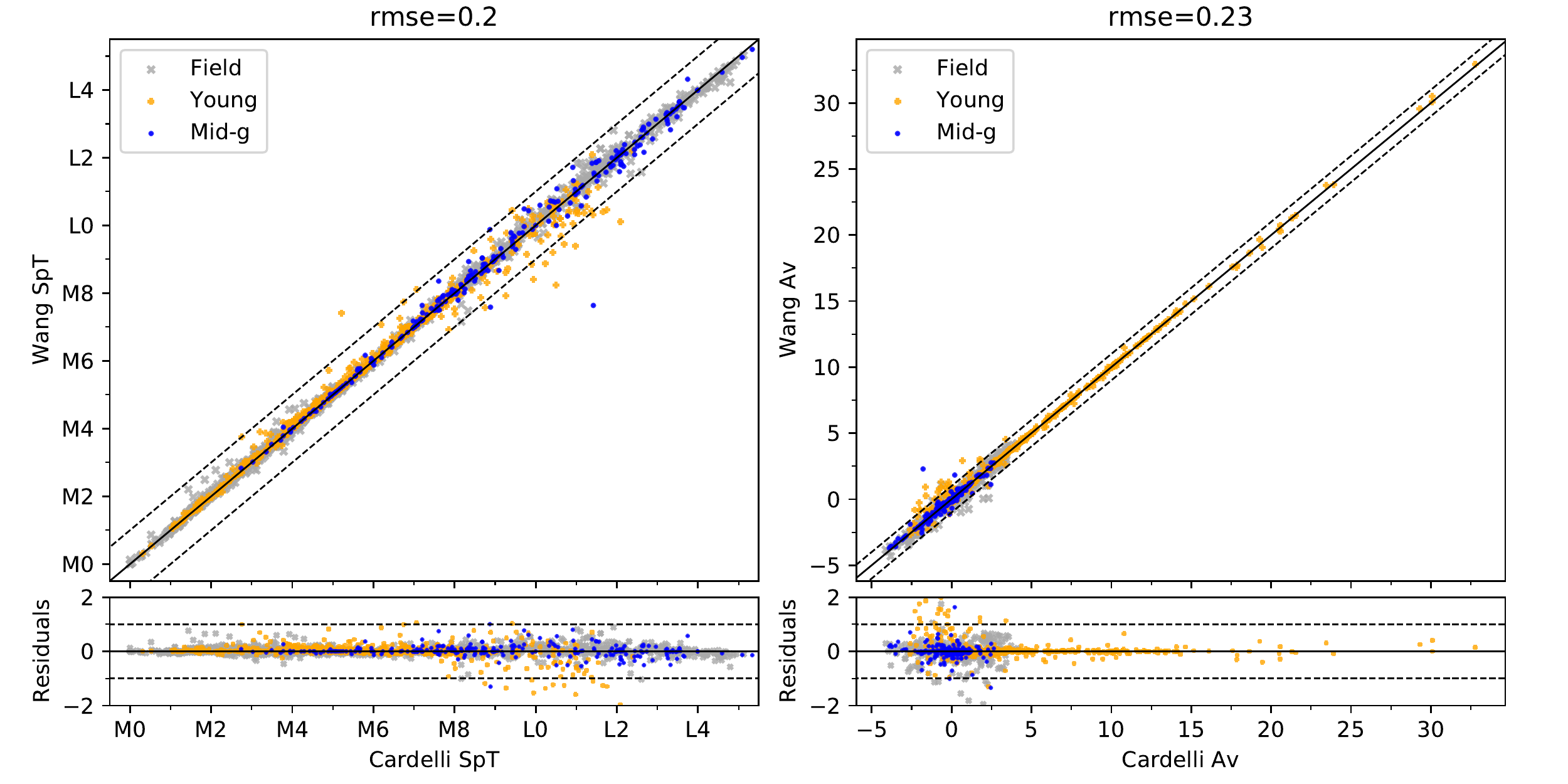}
    \includegraphics[page=2, width=\textwidth]{comp_extinction_law.pdf}
    \caption{Spectral type (left panels) and extinction (right panels) derived from comparison with spectral templates using Cardelli's law compared with the same parameters retrieved using the extinction law from \citet{fitzpatrick99} (top panels) and \citet{wang19} (bottom panels).}
    \label{fig:appendix_ext_law}
\end{figure*}

In Figure~\ref{fig:appendix_ext_law} we show the results of the template fitting process using the extinction law from \citet{cardelli89} compared with the values derived using the extinction law from \citet{fitzpatrick99} and \citet{wang19}. We observe that all the results agree very well, and that the extinction law selected is not affecting the results importantly.

\section{Spectral type indices} \label{appendix_sptindices}

\begin{figure*}[hbt!]
    \centering
    \includegraphics[width=0.95\textwidth]{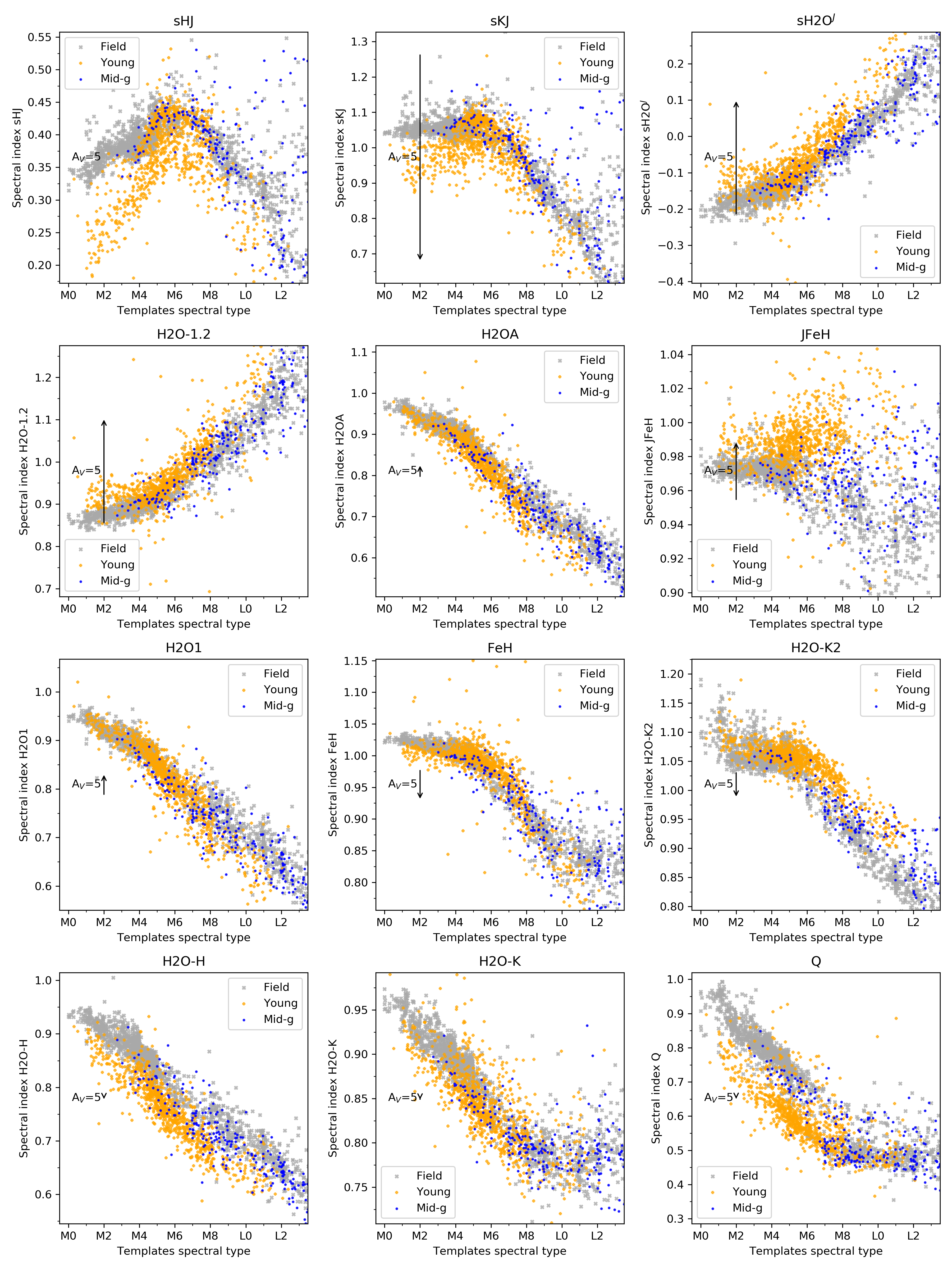}
    \caption{Performance of the non-selected spectral type indices with SpT.}
    \label{fig:appendix1}
\end{figure*}

\begin{figure*}[hbt!]
    \centering
    \includegraphics[width=0.95\textwidth]{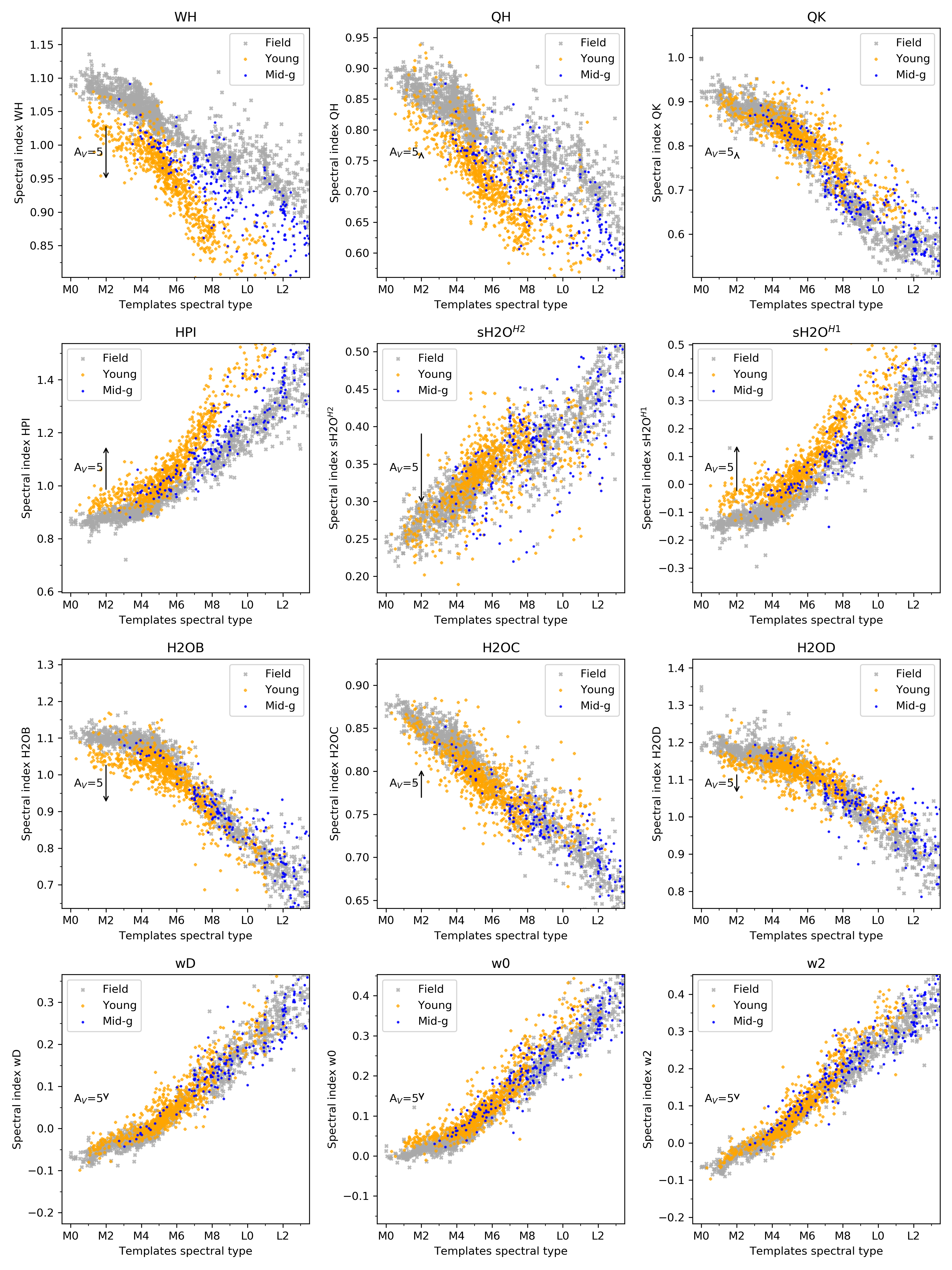}
    \caption{Performance of the non-selected spectral type indices with SpT.}
    \label{fig:appendix2}
\end{figure*}

In Table~\ref{tab:indices_app} we show the functional form of all te indices inspected in section~3.2. In Figures~\ref{fig:appendix1} and \ref{fig:appendix2} we show the calibration with SpT of all the SpT indices that were not selected for further analysis. Some of these indices have been used extensively, most of them were developed using samples with the same characteristics (instrument, spectral resolution, signal to noise ratio and data reduction). On the other hand, our dataset has a heterogeneous nature (which can be seen as a more realistic approach), and it is also much larger than any sample used for the definition of SpT indices. The great size of our dataset allowed us to extract more information on the intrinsic dispersion on the indices, and be able to select the best ones. The baseline spectral typing method is also different from the original works where the indices were defined, which may also play a role in the difference in the behaviour seen here. Additionally, in this work we selected the indices with the best behaviour within our selected SpT range of interest, this may not be true if the range of interest is different. Several of these indices were discarded because they presented a gravity-sensitive behaviour, but some of them show a very good overall performance if used with field objects alone, such are the sH2O$^J$, H2O-K2, HPI and sH2O$^{H1}$ indices.

\renewcommand{\arraystretch}{2}
\begin{table*}
    \caption{Index and reference of the spectral indices shown in Figures \ref{fig:appendix1} and \ref{fig:appendix2}.}
	\begin{center}
        \begin{tabular}{l c c}       
            \hline\hline
            Index & Reference & Formula \\
            \hline
            sHJ & Testi et al. 2001 & $\frac{\langle F_{\lambda=1.265-1.305}\rangle - \langle F_{\lambda=1.6-1.7} \rangle}{0.5(\langle F_{\lambda=1.265-1.305} \rangle + \langle F_{\lambda=1.6-1.7} \rangle )} $ \\
            sKJ & Testi et al. 2001 & $\frac{\langle F_{\lambda=1.265-1.305}\rangle - \langle F_{\lambda=2.12-2.16} \rangle}{0.5(\langle F_{\lambda=1.265-1.305} \rangle + \langle F_{\lambda=2.12-2.16} \rangle )} $ \\
            sH2O$^J$ & Testi et al. 2001 & $\frac{\langle F_{\lambda=1.265-1.305}\rangle - \langle F_{\lambda=1.09-1.13} \rangle}{0.5(\langle F_{\lambda=1.265-1.305} \rangle + \langle F_{\lambda=1.09-1.13} \rangle )} $ \\
            sH2O$^{H1}$ & Testi et al. 2001 & $\frac{\langle F_{\lambda=1.6-1.7}\rangle - \langle F_{\lambda=1.45-1.48} \rangle}{0.5(\langle F_{\lambda=1.6-1.7} \rangle + \langle F_{\lambda=1.45-1.48} \rangle )} $ \\
            sH2O$^{H2}$ & Testi et al. 2001 & $\frac{\langle F_{\lambda=1.6-1.7}\rangle - \langle F_{\lambda=1.77-1.81} \rangle}{0.5(\langle F_{\lambda=1.6-1.7} \rangle + \langle F_{\lambda=1.77-1.81} \rangle )} $ \\
            sH2O$^{K}$ & Testi et al. 2001 & $\frac{\langle F_{\lambda=2.12-2.16}\rangle - \langle F_{\lambda=1.96-1.99} \rangle}{0.5(\langle F_{\lambda=2.21-2.16} \rangle + \langle F_{\lambda=1.96-1.99} \rangle )} $ \\
            H2O-1.2 & Geballe et al. 2002 & $\frac{\langle F_{\lambda=1.26-1.29}\rangle}{\langle F_{\lambda=1.13-1.16} \rangle} $\\
            J-FeH & Mclean et al. 2003 & $\frac{\langle F_{\lambda=1.195-1.205}\rangle}{\langle F_{\lambda=1.18-1.19} \rangle} $\\
            H2OA & Mclean et al. 2003 & $\frac{\langle F_{\lambda=1.338-1.348}\rangle}{\langle F_{\lambda=1.308-1.318} \rangle} $ \\
            H2OB & Mclean et al. 2003 & $\frac{\langle F_{\lambda=1.451-1.461}\rangle}{\langle F_{\lambda=1.565-1.575} \rangle} $\\
            H2OC & Mclean et al. 2003 & $\frac{\langle F_{\lambda=1.783-1.793}\rangle}{\langle F_{\lambda=1.717-1.727} \rangle} $\\
            H2OD & Mclean et al. 2003 & $\frac{\langle F_{\lambda=1.951-1.977}\rangle}{\langle F_{\lambda=2.062-2.088} \rangle} $\\
            H2O-1 & Slesnick et al. 2004 & $\frac{\langle F_{\lambda=1.335-345}\rangle}{\langle F_{\lambda=1.295-1.305} \rangle} $\\
            H2O-2 & Slesnick et al. 2004 & $\frac{\langle F_{\lambda=2.035-2.045}\rangle}{\langle F_{\lambda=2.135-2.145} \rangle} $\\
            FeH & Slesnick et al. 2004 & $\frac{\langle F_{\lambda=1.1935-1.2065}\rangle}{\langle F_{\lambda=1.2235-1.2365} \rangle} $\\
            H2O-K2 & Rojas-Ayala et al. 2012 & $\frac{\langle F_{\lambda=2.07-2.09}\rangle / \langle F_{\lambda=2.235-2.255}\rangle}{\langle F_{\lambda=2.235-2.255}\rangle / \langle F_{\lambda=2.36-2.38}\rangle} $\\
            H2O-H & Covey et al. 2010 & $\frac{\langle F_{\lambda=1.595-1615}\rangle / \langle F_{\lambda=1.68-1.7}\rangle}{\langle F_{\lambda=1.68-1.7}\rangle / \langle F_{\lambda=1.76-1.78}\rangle} $\\
            H2O-K & Covey et al. 2010 & $\frac{\langle F_{\lambda=2.18-2.2}\rangle / \langle F_{\lambda=2.27-2.29}\rangle}{\langle F_{\lambda=2.27-2.29}\rangle / \langle F_{\lambda=2.36-2.38}\rangle} $\\
            Q & Cushing et al. 2000 & $\frac{\langle F_{\lambda=2.07-2.13}\rangle} {\langle F_{\lambda=2.267-2.285} \rangle} \left[ \frac{\langle F_{\lambda=2.4-2.5}\rangle}{\langle F_{\lambda=2.267-2.285} \rangle} \right] ^{1.22} $\\
            WH & Weights et al. 2009 & $\frac{\langle F_{\lambda=1.552-1.572}\rangle}{\langle F_{\lambda=1.655-1.675} \rangle} $\\
            WK & Weights et al. 2009 & $\frac{\langle F_{\lambda=2.04-2.06}\rangle}{\langle F_{\lambda=2.18-2.2} \rangle} $\\
            QH & Weights et al. 2009 & $\frac{\langle F_{\lambda=1.552-1.572}\rangle}{\langle F_{\lambda=1.655-1.675} \rangle} \left[ \frac{\langle F_{\lambda=1.73-1.75}\rangle}{\langle F_{\lambda=1.655-1.675} \rangle} \right] ^{1.581} $\\
            QK & Weights et al. 2009 & $\frac{\langle F_{\lambda=2.04-2.06}\rangle}{\langle F_{\lambda=2.182-2.202} \rangle} \left[ \frac{\langle F_{\lambda=2.33-2.35}\rangle}{\langle F_{\lambda=2.182-2.202} \rangle} \right] ^{1.14} $\\
            HPI & Scholz et al. 2012 & $\frac{\langle F_{\lambda=1.675-1.685}\rangle}{\langle F_{\lambda=1.495-1.505} \rangle} $\\
            wO & Zhang et al. 2018 & $\left( \frac{W_{H2O}}{-0.0105}^*-\frac{W_{H2O-1}}{-0.0102}^* \right) \left( \frac{1}{-0.0105}+\frac{1}{-0.0102} \right)^{-1}$ \\
            wD & Zhang et al. 2018 & $\left( \frac{W_{H2O-D}}{0.0099}^*-\frac{W_{H2O-1}}{-0.0102} \right) \left( \frac{1}{0.0099}+\frac{1}{-0.0102} \right)^{-1}$ \\
            w2 & Zhang et al. 2018 & $\left( \frac{W_{H2O-2}}{0.0098}^*-\frac{W_{H2O-1}}{-0.0102} \right) \left( \frac{1}{0.0098}+\frac{1}{-0.0102} \right)^{-1}$ \\
            \hline 
		\end{tabular}
		\tablefoot{
		$^*$ $W_{index}=-2.5\log{(index)}$, where index are the H2O \citep{allers07}, H2O-D \citep{mclean03}, H2O-2 and H2O-1 \citep{slesnick04} spectral indices.
        }
	\end{center}
	\label{tab:indices_app}
\end{table*}
\renewcommand{\arraystretch}{1}

\section{Machine learning} \label{appendix_ml}

\subsection{Machine learning models}

We compare the performance of four different models. The first two are applied to the previously computed TLI-g index and SpT, while we use the entire spectra as input for the latter two methods. We adopted the following models:
\begin{enumerate}
    \item \textbf{LDA:} We apply Linear Discriminant Analysis \citep[LDA;][]{FISHERLDA,RAOLDA} to the TLI-g index vs. SpT plane. LDA is a linear classifier that performs classification using Bayes probability assuming a normal distribution of the variables.
    \item \textbf{SVM:} We apply Support Vector Machine \citep[SVM;][]{SVMBoser, SVMCortes} to the TLI-g index and SpT. SVM maximizes the distance between the different classes and a separating hyperplane defined on a feature space that is transformed using a kernel.
    \item \textbf{PCA+SVM:} Here, we first perform feature extraction from the entire spectra by importance/variance with Principal Component Analysis \citep[PCA;][]{PCAPearson,PCAHotelling}. The original features are the entire wavelength grid. Then we apply SVM for classification on the projection of the spectra in the principal components that explain the most variance.
    \item \textbf{RF:} Random Forest \citep[RF;][]{RFHo,RFBreiman} is an ensemble of regression trees that implements bagging (bootstrapping with aggregation). We apply RFs directly to the entire spectra, without any kind of feature extraction.
\end{enumerate}

\subsection{Selection of hyper-parameters}

In the SVM we used a standard Radial Basis Function ($rbf$) kernel. This kernel has the form $K(\mathbf{x},\mathbf{x'})=\exp{(-\gamma||\mathbf{x}-\mathbf{x'}||^2)}$, where $\mathbf{x}$ and $\mathbf{x'}$ are pairs of feature vectors and $\gamma$ is a hyper-parameter. In total, SVM has two hyper-parameters that require optimization:
\begin{itemize}
    \item Penalty parameter C: This parameter accounts for the number and severity of the violations to the hyperplane generated by the SVM. A very low C value means that the model highly fits the data, being more prone to overfitting, and as the value gets higher the fitting is less hard.
    \item $\gamma$: It is a parameter of the $rbf$ kernel as defined above. It can be seen as inversely proportional to the variance of a gaussian, i.e. $\gamma=1/(2\sigma^2)$, and acts as a regularization parameter together with C, therefore they need to be optimized together.
\end{itemize}

We performed the optimization of the hyper-parameters of the SVM and PCA+SVM methods  via grid search in $\gamma$ and C and number of components for PCA+SVM. For each grid position we derive the mean precision and recall of a 5-fold segmented cross validation repeated 5 times. In each of the 5 repetitions, the dataset is shuffled and randomly divided in 5 parts. The partitions are segmented in order to tackle class imbalance, meaning that within a given partition there is a random 20\% of objects from each class. The model is then trained with 4 of these divisions and tested on the 5th. The process is repeated until the 5 divisions have been used as the test set (segmented cross-validation). Cross-validation gives an insight on the generalization of the model, reducing the selection bias and overfitting that can arise from a simple train/test run. By repeating the cross-validation 5 times, we ensure that the results are minimally affected by the random divisions of the data in the cross-validation process. The hyper-parameters are selected from the region with the highest metrics (precision and recall). We observe that the results are not significantly affected by a change in the hyper-parameters.

In RF, two hyper-parameters affect mostly its performance: the number of trees and the maximum number of features. As the number of trees is increased the performance of RF will improve, but it will eventually reach a stabilization point. We find this point at 200 trees for all the cases we tested. We also observed that the maximum number of features included in each tree could affect the performance of RF. A common rule of thumb that is adopted as the default value of this hyper-parameter is the square root of the number of features. We found that in some cases when this value is increased the performance of the model increased significantly. We evaluated this hyper-parameter for a wide range of values with a fixed number of trees. Both hyper-parameters are selected as in the previous case, from the mean precision and recall of a 5-fold segmented cross validation repeated 5 times.

The selected hyper-parameters are shown in Table~\ref{tab:hyper-parm} for the four cases inspected: predicting for 2 or 3 age classes, and applied to the HK or JHK bands spectra (only applicable for the PCA+SVM and RF models where the entire spectrum is used as input).

\subsection{Performance evaluation}

Among the methods adopted herein, LDA, SVM and PCA need the dataset to be normalized, meaning that the mean and standard deviation of each feature has to be 0 and 1 respectively. Otherwise, the intrinsic variation of each feature will prevent a correct classification.

Once the hyper-parameters have been selected (see Table~\ref{tab:hyper-parm}) we proceed to run the different models. We evaluate the performance of the models based on the precision and recall metrics and the confusion matrix. The confusion matrix allows a more straightforward extraction of information on the performance of the models. We run the models with a segmented 80/20 train/test configuration, and we repeat the process 200 times. Before each iteration the dataset is shuffled and the train/test division is made randomly. In the end, each object will have appeared approximately 10 times in the test set, and will therefore have $\sim$10 different classifications, each associated with a score according to the class that it has been classified into. The confusion matrix is built from the mean predicted age class score of all the objects.

\begin{table}
    \caption{Hyper-parameters selected for each of the models in the four case scenarios inspected. n\_c is the number of components used, n\_trees the number of trees, and max\_f is the maximum number of features included in each tree.}
	\begin{center}
        \begin{tabular}{l c c c c c c c}       
            \hline\hline
            & \multicolumn{2}{c}{SVM} & \multicolumn{3}{c}{PCA+SVM} & \multicolumn{2}{c}{RF} \\
             & $\gamma$ & C & $\gamma$ & C & n\_c & n\_trees & max\_f \\
            \hline
            2-JHK & 0.3 & 40 & 0.012 & 50 & 9 & 200 & 14 \\
            2-HK & - & - & 0.012 & 45 & 12 & 200 & 40 \\
            3-JHK & 0.05 & 30 & 0.012 & 30 & 10 & 200 & 70 \\
            3-HK & - & - & 0.012 & 45 & 12 & 200 & 70 \\
            \hline 
		\end{tabular}
	\end{center}
    \label{tab:hyper-parm}
\end{table}

All the models were implemented within scikitlearn \citep{scikit}.

\subsection{Feature importance}

\begin{figure*}[hbt!]
    \centering
    \includegraphics[width=\textwidth]{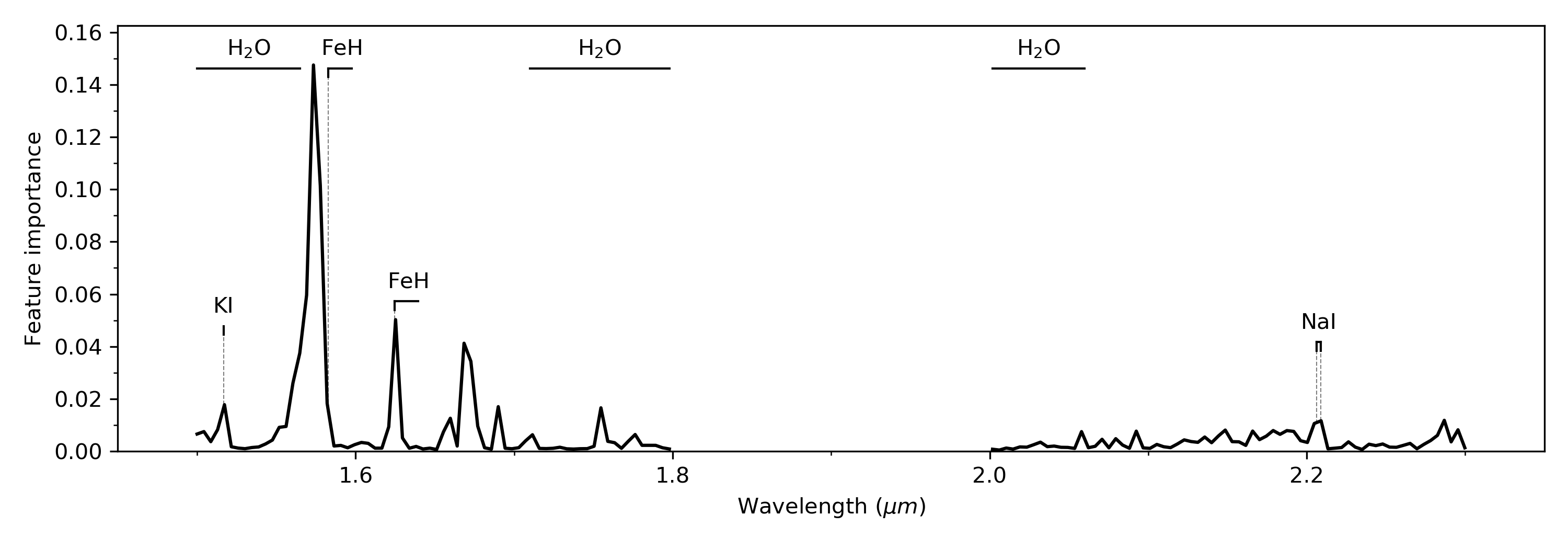}
    \caption{RF feature importance for the HK bands range, when classifying for two age classes: field and young. We also show the known gravity-sensitive features as well as the water bands, the main driver of the cool dwarfs broadband shape.}
    \label{fig:importance_HK_2}
\end{figure*}

\begin{figure*}[hbt!]
    \centering
    \includegraphics[width=\textwidth]{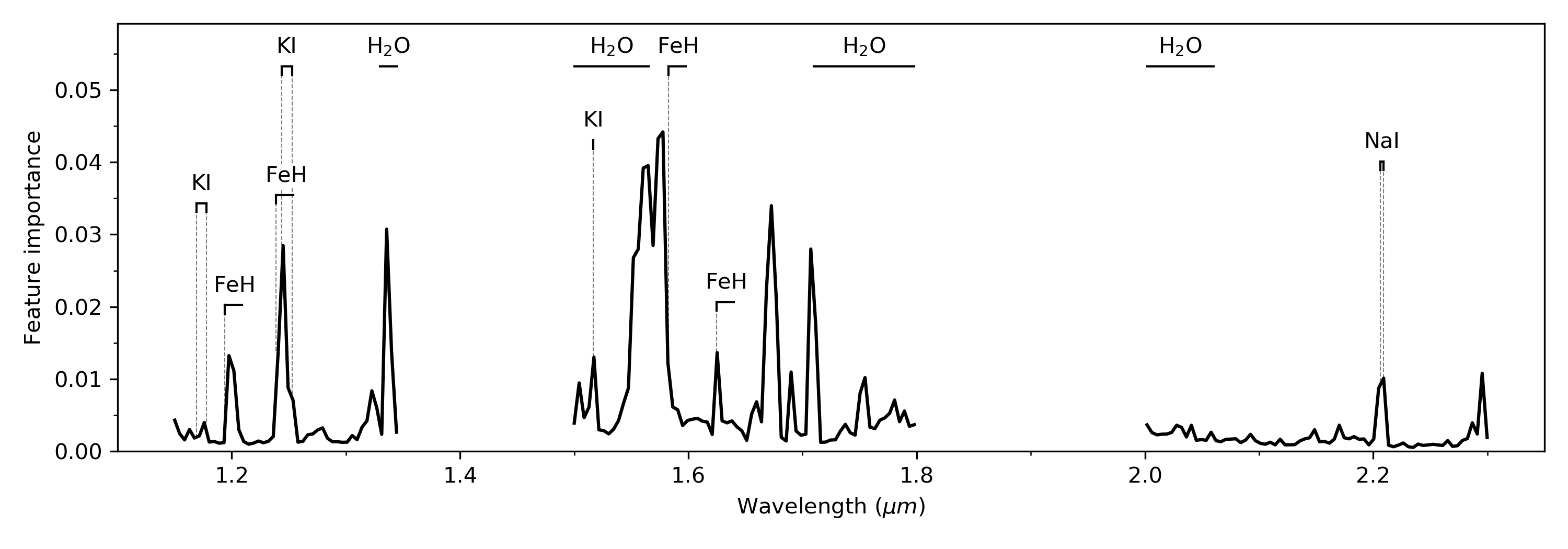}
    \caption{RF feature importance for the entire JHK range, when classifying for three age classes: field, mid-gravity and young. We also show the known gravity-sensitive features as well as the water bands, the main driver of the cool dwarfs broadband shape.}
    \label{fig:importance_JHK_3}
\end{figure*}

\begin{figure*}[hbt!]
    \centering
    \includegraphics[width=\textwidth]{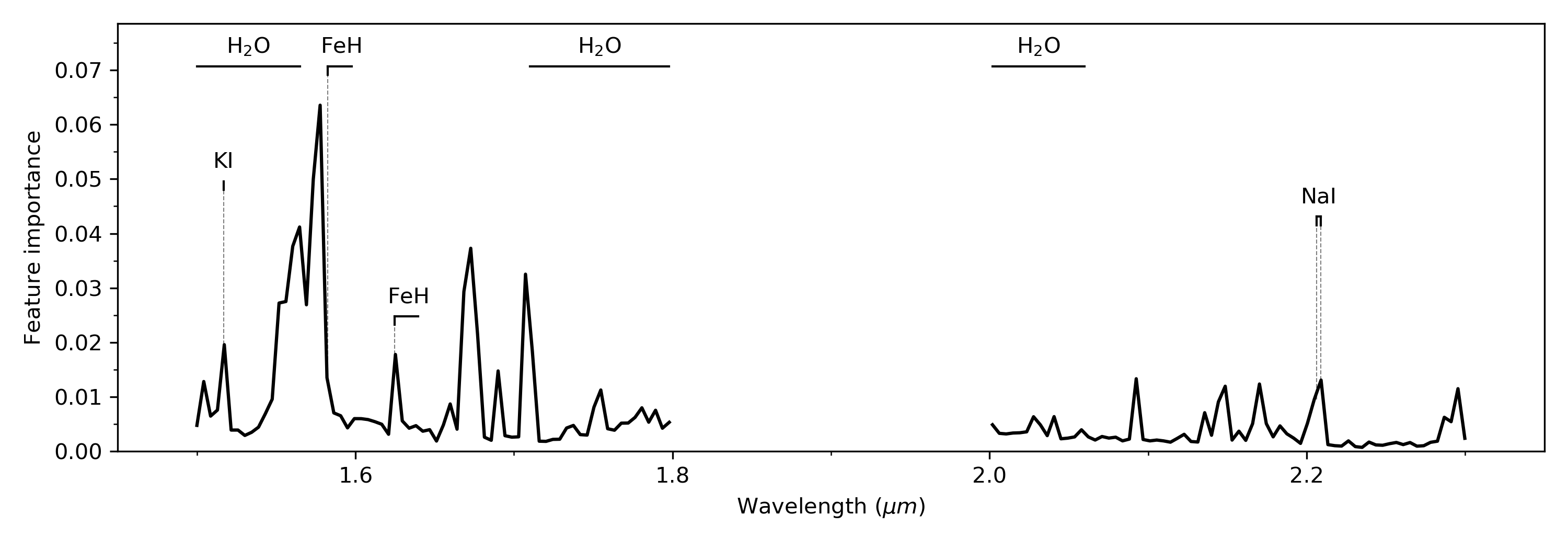}
    \caption{RF feature importance for the HK bands range, when classifying for three age classes: field, mid-gravity and young. We also show the known gravity-sensitive features as well as the water bands, the main driver of the cool dwarfs broadband shape.}
    \label{fig:importance_HK_3}
\end{figure*}

In Figures~\ref{fig:importance_HK_2},~\ref{fig:importance_JHK_3} and \ref{fig:importance_HK_3} we show the feature importance for the cases inspected in Section~\ref{mlearning}.

\end{appendix}


\end{document}